\documentclass[fleqn,usenatbib]{mnras}

\usepackage{newtxtext,newtxmath}

\usepackage[T1]{fontenc}

\DeclareRobustCommand{\VAN}[3]{#2}
\let\VANthebibliography\thebibliography
\def\thebibliography{\DeclareRobustCommand{\VAN}[3]{##3}\VANthebibliography}

\makeatletter
\newlength{\abovecaptionskip}%
\makeatother

\usepackage{amsmath,amstext}
\usepackage{graphicx}	
\usepackage[figure,figure*]{hypcap}
\usepackage{enumerate}
\usepackage{xspace} 
\usepackage{multirow}
\usepackage[flushleft]{threeparttable}
\usepackage{booktabs}
\usepackage{adjustbox}
\usepackage{marginnote}

\usepackage{soul}
\setstcolor{red}

\makeatletter
\newcommand{\HI}{{\text{H\MakeUppercase{\romannumeral 1}}}\xspace}
\newcommand{\HII}{{\text{H\MakeUppercase{\romannumeral 2}}}\xspace}
\newcommand{\HeI}{{\text{He\MakeUppercase{\romannumeral 1}}}\xspace}
\newcommand{\HeII}{{\text{He\MakeUppercase{\romannumeral 2}}}\xspace}
\newcommand{\HeIII}{{\text{He\MakeUppercase{\romannumeral 3}}}\xspace}
\newcommand{\Lya}{\ifmmode{\mathrm{Ly}\alpha}\else Ly$\alpha$\xspace\fi}
\newcommand{\Htwo}{\ifmmode{\mathrm{H}_2}\else H$_2$\xspace\fi}
\newcommand{\OIII}{[{\text{O\MakeUppercase{\romannumeral 3}}}]\xspace}
\newcommand{\threeHplus}{\ensuremath{{}^3{\rm He}^+ \xspace }}

\newcommand{\DI}{{\text{D\MakeUppercase{\romannumeral 1}}}\xspace}
\newcommand{\MBII}{{\text{MB\MakeUppercase{\small\expandafter{\romannumeral 2}}}}\xspace}

\newcommand{\Rmnum}[1]{\expandafter\@slowromancap\romannumeral #1@}

\newcommand{\xHII}{x_{\mbox{\tiny H\Rmnum{2}}}}
\newcommand{\xHeII}{x_{\mbox{\tiny He\Rmnum{2}}}}
\newcommand{\xHeIII}{x_{\mbox{\tiny He\Rmnum{3}}}}

\newcommand{\vT}{\ensuremath{\langle\, T \rangle}}
\newcommand{\vHI}{\ensuremath{\langle x_{\mbox{\tiny H\Rmnum{1}}} \rangle}}
\newcommand{\vHII}{\ensuremath{\langle x_{\mbox{\tiny H\Rmnum{2}}} \rangle}}

\newcommand{\vHeII}{\ensuremath{\langle x_{\mbox{\tiny He\Rmnum{2}}} \rangle}}
\newcommand{\vHeIII}{\ensuremath{\langle x_{\mbox{\tiny He\Rmnum{3}}} \rangle}}
\newcommand{\vun}{\ensuremath{\langle u_0 \rangle}}

\newcommand{\Msun}[1]{\ensuremath{{\rm M}_{\odot}} #1 }
\newcommand{\h}[1]{\ensuremath{h^{-1}} #1 }
\newcommand{\diff}{\ensuremath{\; \text{d}}}

\newcommand{\abb}[2]{{#1}{\small \uppercase\expandafter{\romannumeral #2}}}
\newcommand{\abbm}[2]{\ensuremath{{\rm #1}{\small\rm\scriptstyle \uppercase\expandafter{\romannumeral #2}}}}

\title[The Epoch of Reionization]{Large scale simulations of H and He reionization and heating driven by stars and more energetic sources}

\author[M.B.~Eide et al.]{Marius B. Eide$^{1}$\thanks{E-mail: \href{mailto:eide@MPA-Garching.MPG.DE}{eide@MPA-Garching.MPG.DE}}
Benedetta Ciardi$^{1}$, 
Luca Graziani$^{2,3,4}$,
Philipp Busch$^{1}$,
Yu Feng$^{5}$
\newauthor
and Tiziana Di Matteo$^{6}$
\\
$^{1}$Max-Planck-Institut f\"ur Astrophysik, Karl-Schwarzschild-Stra\ss e 1, D-85741 Garching, Germany\\
$^{2}$Dipartimento di Fisica, Sapienza, Universit$\grave{a}$ di Roma, Piazzale Aldo Moro 5, 00185, Roma, Italy\\
$^{3}$INAF/Osservatorio Astrofisico di Arcetri, Largo E. Fermi 5, 50125 Firenze, Italy\\
$^{4}$INFN, Sezione di Roma I, P.le Aldo Moro 2, 00185 Roma, Italy\\
$^{5}$Berkeley Center for Cosmological Physics Campbell Hall 341, University of California, Berkeley CA 94720, United States\\
$^{6}$McWilliams Center for Cosmology, Physics Department, Carnegie Mellon University, Pittsburgh, PA 15213, USA
}

\date{Accepted XXX. Received YYY; in original form ZZZ}

\pubyear{2020}

\begin{document}
\label{firstpage}
\pagerange{\pageref{firstpage}--\pageref{lastpage}}
\maketitle

\begin{abstract}
We present simulations of cosmic reionization and reheating from $z=18$ to $z=5$, investigating the role of stars (emitting soft UV-photons), nuclear black holes (BHs, with power-law spectra), X-ray binaries (XRBs, with hard X-ray dominated spectra), and the supernova-associated thermal bremsstrahlung of the diffuse interstellar medium (ISM, with soft X-ray spectra). We post-process the hydrodynamical simulation Massive-Black II (MBII) with multifrequency ionizing radiative transfer. The source properties are directly derived from the physical environment of MBII, and our only real free parameter is the ionizing escape fraction $f_{\rm esc}$. We find that, among the models explored here, the one with an escape fraction that decreases with decreasing redshift yields results most in line with observations, such as of the neutral hydrogen fraction and the Thomson scattering optical depth. Stars are the main driver of hydrogen reionization and consequently of the thermal history of the intergalactic medium (IGM). We obtain $\vHII = 0.99998$ at $z=6$ for all source types, with volume averaged temperatures $\vT \sim 20,000$~K. BHs are rare and negligible to hydrogen reionization, but conversely they are the only sources which can fully ionize helium, increasing local temperatures by $\sim 10^4$~K. The thermal and ionization state of the neutral and lowly ionized hydrogen differs significantly with different source combinations, with ISM and (to a lesser extent) XRBs, playing a significant role and, as a consequence, determining the transition from absorption to emission of the 21~cm signal from neutral hydrogen.  
\end{abstract}


\section{Introduction}
\label{sec:intro}

The existence and characteristics of the last phase change of our Universe---the Epoch of Reionization (EoR, see e.g.~\citealt{Zaroubi2013} for a review) has been a puzzle ever since the cosmological models of last century envisioned some sort of matter, likely hydrogen, to exist in between galaxies \citep[an intergalactic medium,][]{Hoyle1948}, which thus could appear as an absorption feature in the spectra of high redshift quasars \citep{Davies1964,Gunn1965}. Another half a century of search \citep[e.g][]{Penzias1968,Field1972,Schneider1991} has not only ultimately uncovered the IGM's existence \cite[e.g.~][]{Fan2006}, but also coincidentally revealed the afterglow from the primordial cosmic fireball \citep[CMB,][]{Penzias1965}. Various probes (e.g.~the disappearance of Lyman $\alpha$ radiation at higher redshifts, \citealt{Dijkstra2014}; CMB scattering off electrons freed in the EoR, \citealt{PlanckCollaborationVI2018}; absorption in the spectra of gamma ray bursts and QSOs, \citealt{Ciardi2005}; as well as the intrinsic 21~cm glow of the neutral IGM, \citealt{Madau1997}) indicate that the IGM transition from neutral to ionized occurred more than 12 billion years ago. The details of its occurrence, however, remain an outstanding and heavily investigated question.

Stellar type sources are believed to be the dominant driver of the EoR \citep{Robertson2015} if a sufficiently large fraction of the ionizing stellar radiation generated in galaxies escapes \citep[e.g.~][]{Vanzella2018,Matthee2018,Steidel2018}.
Nuclear black holes (BHs) are long believed to play a crucial role, particularly in recent years after the discovery of a population of faint quasars (QSOs; \citealt{Giallongo2015}), indications of an extended and patchy reionization of helium (\citealt{Worseck2016,Worseck2018}) and enormous, opaque troughs in an otherwise partially ionized IGM \citep{Becker2015,Barnett2017} that fail to be explained with stellar sources alone \citep{Chardin2015,Chardin2017}. Estimating the BHs' role requires sampling of the faint end of their luminosity function at high $z$ \citep[e.g.~][]{Parsa2018,Wang2018,Kulkarni2018}. 
An additional source that has been somewhat overlooked, but nevertheless of potential importance, is the thermal Bremsstrahlung from the interstellar medium (ISM), driven by the cooling of supernova-shocked gas. It has been observed as a diffuse X-ray component in galaxies \citep[e.g.~][]{Mineo2012a}, and its existence may be followed by intergalactic heating \citep{Meiksin2017,Eide2018}. Its softer spectrum potentially likens its signatures to those of the more exotic cosmic rays \citep[e.g.~][]{Sazonov2015,Leite2017,Jana2018} and AGB stars \citep{Vasiliev2018}.
Finally, high- and low-mass X-ray binaries (XRBs) may also play a role. They emanate harder radiation from accretion disks in stellar systems where a star is devoured by a BH, a neutron star or potentially a white dwarf. While XRBs dominate the X-ray output of gas-poor galaxies \citep{Fabbiano2006,Mineo2012}, their hard spectra are not necessarily primarily accompanied by IGM ionization, but rather by heating \citep{Fialkov2014}.

As already discussed by \cite{Eide2018} in the context of Cosmic Dawn, there is a subtle interplay between the different sources of ionizing radiation, which is particularly relevant for a correct determination of the IGM temperature, but also for partially ionized H and He ionization. This, in turn, is expected to have an important impact, among others, on the modeling of the 21~cm signal from neutral hydrogen (see e.g. \citealt{CiardiMadau2003}). Here, we thus push the simulations presented in \cite{Eide2018} to a lower redshift and concentrate on the analysis of the effect of different sources of ionizing radiation at a later time, for $z \lesssim 10$.

To model the EoR including a variety of sources, we post-process outputs of the hydrodynamic simulation MassiveBlack-II \citep[MBII,][]{Khandai2015} with the 3D multifrequency radiative transfer (RT) code \texttt{CRASH} \citep{Ciardi2001,Maselli2003,Maselli2009,Graziani2013,Hariharan2017,Graziani2018,Glatzle2019}.
Our approach to modelling the EoR differs slightly from models where sources are populated solely based on the dark matter halo mass \citep[e.g.~][]{iliev2006,McQuinn2007,Trac2007,Ross2017}, but it is similar in its multifrequency and post-processing nature (see also \citealt{Baek2010,Paardekooper2013}). In comparison to monofrequency approaches in which the rays' properties (and impact) do not change as they propagate through the IGM, a multifrequency RT allows to better capture the impact of radiation on the ionization and thermal state of the IGM \citep[e.g.~][]{Ciardi2012}. It should be noted that as our RT is not fully coupled to the cosmological simulations \citep[e.g.~][]{Gnedin2000,Gnedin2014,Norman2015,Ocvirk2016,Pawlik2017,Katz2018UVB,Rosdahl2018,Finlator2018}, the ionization and thermal state do not provide feedback and affect subsequent galaxy formation. Our approach, though, is computationally cheaper and thus allows for a more accurate modeling of the RT (e.g. highly resolved spectra of the sources and inclusion of both H and He) and of the ionization and thermal state of the IGM.
For completeness, we also mention another category of approaches employed to model cosmic reionization, namely semi-numerical methods (e.g. \citealt{MesingerFurlanetto2007,Choudhury2009}), which are typically based on an excursion-set approach and thus particularly fast and ideal when a large parameter exploration is required at the expense of a more accurate determination of the physical properties of the IGM. In this respect a good compromise is achieved by methods based on a combination of N-body simulations and post-processing with 1D (rather than 3D) RT codes (e.g. \citealt{Thomas2009,Ghara2018}).

Our work is structured as follows.
In \autoref{sec:methods} we present the details of \MBII, the RT and the source modeling.
In \autoref{sec:results} we present our results in terms of discussion of the physical state of the IGM as determined by different source types, as well as of comparison to available observations.
We discuss the results and give our conclusions in \autoref{sec:discussion}.

\section{Method}
\label{sec:methods}

The approach followed to model cosmic reionization is the same one presented in \citet[][in the following Eide2018]{Eide2018}. Here we outline its main characteristics, and refer the reader to the original paper for more details.

To retrieve the properties of the environment and of the sources of ionizing photons we use the cosmological SPH simulation MassiveBlack-II \citep[MBII,][]{Khandai2015} with sides of length $100\h$ cMpc\footnote{The cosmological parameters adopted in the simulation are taken fom WMAP7 \citep{Komatsu2011}: $\sigma_8 = 0.816$, $n_s = 0.968$, $\Omega_\Lambda = 0.725$, $\Omega_{\rm m} = 0.275$, $\Omega_{\rm b} = 0.046$ and $h=0.701$.}. 
It was run with \texttt{P-GADGET} \citep[see][for details concerning the earlier \texttt{GADGET-2}]{Springel2005a} using $2 \times 1792^3$ dark matter and gas particles of mass $m_{\rm DM} = 1.1\times 10^7 \h \Msun$ and $m_{\rm gas} = 2.2 \times 10^6 \h \Msun$, respectively. The transition of gas into stars is governed by the sub-grid physics presented in \cite{Springel2003} (a comparison to observed star formation rates can be found in Fig.~6 of \citealt{Khandai2012} and in Fig.~23 of \citealt{Khandai2015}). Black holes form following the prescriptions by \cite{DiMatteo2005} and \cite{Springel2005a}. 
MBII also accounts for the resulting feedback processes \citep{DiMatteo2008,Croft2009,Degraf2010,DiMatteo2012}.
We grid the 15 snapshots\footnote{The redshift of the snapshots is dictated by the outputs of the MBII simulations and it is $z$=18, 16, 14, 13, 12, 11, 10, 9, 8, 7.5, 7, 6.5, 6, 5.5, 5.} between $z=18$ and $5$ onto $256^3$ cells, resulting in a resolution of $391\h$ ckpc.
This gives us locations of potential ionizing sources through stellar particles, halos and black holes representative, respectively, of stellar populations, galaxies and galactic nuclei in various states of activity. We also grid the gas temperature and density\footnote{The gas density is translated to hydrogen and helium number densities assuming number fractions of $X=0.92$ and $Y=0.08$, respectively, and no metals.}. When multiple source candidates appear within the same cell, we add together their derived spectral shapes and luminosities. 

At $z<10$ we further reduce the number of sources by evenly absorbing the luminosity of faint cells into all their neighbours that are at least $100$ times brighter, if they themselves are not at least 100 times brighter than any neighbour. This procedure is iterated until no more cells can be absorbed into brighter neighbours. The result is a source field with a greatly reduced number of sources. The structure of this new field still closely resembles that of the original one, as only very faint extensions of larger source clusters are integrated into the brighter central cells that would outshine them in any case. Isolated faint sources are left untouched. 
The combination reduces the number of source by 50\% at $z=10$ and up to 70\% at $z=7$.
Note that, with this procedure, at $z=7$ the vast majority of sources are displaced by less than 1~$h^{-1}$cMpc and virtually all remain within 2~$h^{-1}$cMpc, i.e. much less than the typical dimension of an ionized region at the same redshift \citep[see also~][]{Busch2020}.
As a reference, for the simulations with all sources included and $f_{\rm esc}=15\%$ at $z=8.56$ we find that clustering the sources induces a difference in the volume averaged temperature and HII and HeII fractions of 0.09\%, which increases to 0.3\% for the HeIII fraction. These values are within the numerical noise as they are below our convergence limits. A further comparison of cell by cell values indicates that the clustering of the sources does not affect the results presented in the paper.

\begin{figure}
\includegraphics[width=\columnwidth]{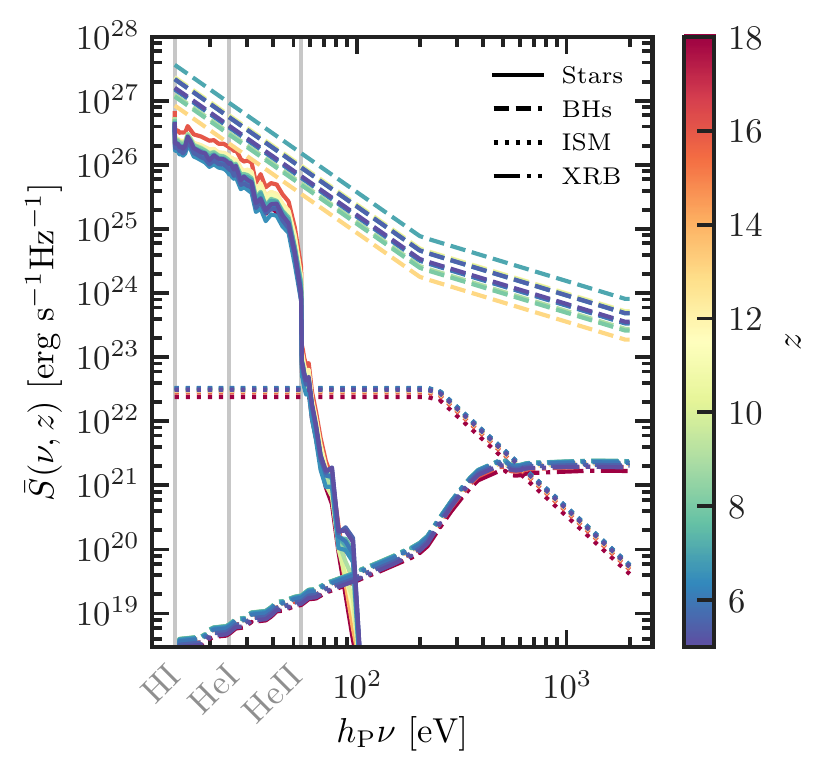}
\caption{Average SEDs for each source type, stars (solid lines), BHs (dashed), ISM (dotted) and XRBs (dash dot-dotted) as function of photon energy plotted at various redshifts (indicated by the line colour). The ionization thresholds for \HI, \HeI and \HeII are plotted as vertical grey lines.}
\label{fig:SEDs}
\end{figure}

Next, we detail how  the spectral and luminous properties of the sources are obtained.
For each source $i$ at a redshift $z$, we calculate the ionizing emissivity between $13.6$ eV and $2$ keV as
\begin{equation}
\varepsilon_i (z) = \int\limits_{13.6\,\rm eV}^{2\,\rm keV} \frac{S_i(\nu,z)}{h^2_{\rm P} \nu} \diff \left(h_{\rm P}\nu \right),
\end{equation}
where $h_{\rm P}$ is the Planck's constant, $\nu$ is the frequency in Hz, and $S_i$ (in erg s$^{-1}$ Hz$^{-1}$) is the source's spectrum, which is evaluated following the procedure described in Eide2018. 
As in Eide2018 the emissivities of stars, XRBs and ISM sharing the same cell are summed together and assigned a spectrum which is the sum of the volume averaged spectra of each source type. BHs are instead treated separately, as not every galaxy hosts an active BH. In Fig.~\ref{fig:SEDs} we show the spectral energy distribution (SED) of the various sources at different redshifts. Note that this is very similar to Fig.~2 of Eide2018, but has been nevertheless included to show the spectra at all redshift for completeness. As discussed in Eide2018, the evolution in $z$ is extremely mild. The SEDs of the various sources are discussed in the following.

\begin{figure}
\includegraphics[width=\columnwidth]{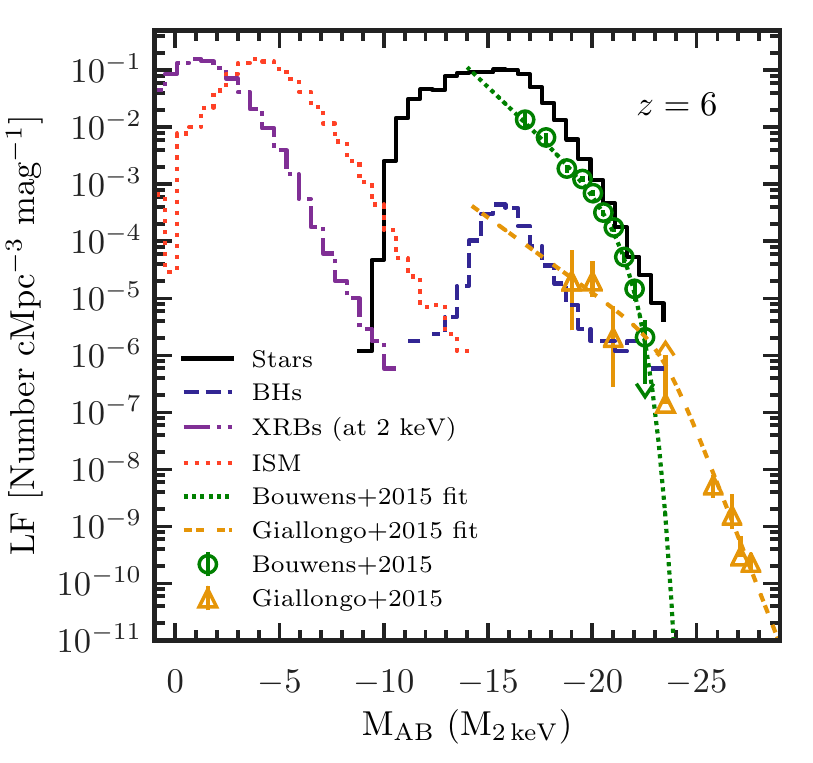}
\caption{Luminosity functions at AB magnitude 1450~\AA~at $z=6$ for stars (solid black line), BHs (dashed blue), ISM (dotted red), and at 6.2~\AA (i.e. 2~keV) for XRBs (dash dot-dotted purple). The green circles and yellow triangles are observational data for stars \citep{Bouwens2015LF} and BHs \citep{Giallongo2015}, respectively, while the dotted and double-dashed lines are corresponding fits.}
\label{fig:LF}
\end{figure}

In Fig.~\ref{fig:LF} we show the luminosity function (LF) of the various source types at $z=6$, estimated at an absolute magnitude ${\rm M}_{\rm AB}$ at $1450$~\AA\footnote{Except for XRBs, for which we estimate the LF at a wavelength of $6.2$~\AA, corresponding to photon energies of $2$~keV.}. We note that the LF is calculated before applying any grid shifting of the sources and estimated on a $1024^3$ grid, i.e.~with sources grouped together in cells with widths of 143 ckpc, slightly larger than present-day Milky Way-sizes.
The LFs are discussed in the following, where we summarize the main features of the four different adopted source types.


\textit{Stars}: The spectra of stars are obtained from the 2012 version of the stellar population synthesis code \texttt{BPASS} \citep{Eldridge2012} by evaluating a spectrum for each star particle given its stellar mass, age and metallicity. We do not account for contributions from stars that have evolved into binary systems, i.e.~we use the `single star' version of BPASS, nor for nebular emission.
As a reference, at $z=6$ we have a total of $3.1 \times 10^{7}$ star particles. The volume averages of the individual galaxy spectra are strongest at \HI and \HeI ionizing frequencies, and fall by orders of magnitude near the \HeII ionization threshold, as seen in Fig.~\ref{fig:SEDs}. The spectra display little redshift evolution towards lower $z$ except for a dampening of the spectra at lower energies and a hardening at energies $> 54.4$ eV.

The LF has a sharp cut-off at both faint (${\rm M}_{\rm AB} > -10$) and bright (${\rm M}_{\rm AB} < -18$) magnitudes, with a plateau and peak near ${\rm M}_{\rm AB} \sim -15$.
There is a sizeable contribution from faint stellar sources, as well as a population of sources that are brighter than most BHs, except for the very brightest BHs at ${\rm M}_{\rm AB} \sim -23$. The LF from the simulation matches observational constraints well \citep{Bouwens2015LF}, showing a similar distribution of brighter stars and reproducing the sharply rising behaviour of a Schechter LF. At the fainter magnitudes, where no observational constraint is available, the LF decreases, a feature shown to be related to a lower SFR and stellar content in smaller mass halos (see e.g. \citealt{OShea2015}). A reduction in SFR in low-mass galaxies due to photoionization feedback has also been found in fully coupled reionization simulations \citep{Dawoodbhoy2018}.

\textit{Nuclear Black Holes (BHs)}: In MBII each BHs is modelled as a collisionless sink particle of mass $5 \times 10^5 \, h^{-1}$~M$_\odot$ seeded within newly formed halos with mass above $5 \times 10^{10} \, h^{-1}$~M$_\odot$. The BH seed grows by accreting surrounding gas or by merging with other BHs. The accretion rate, $\dot{M}_i$, depends on the BH mass, its velocity relative to the surrounding gas and the density and sound speed of the gas. $\dot{M}_i$ is mildly super-Eddington.
Each BH in MBII is assumed to radiate with a bolometric luminosity $L^{\rm BH}_i = \eta \dot{M}_i c^2$  (in erg s$^{-1}$) given by its accretion rate $\dot{M}_i$ (in g s$^{-1}$) scaled by an efficiency parameter $\eta$ and the speed of light $c$ \citep{Shakura1973}. The efficiency parameter $\eta = 0.1$ is chosen to be consistent with MBII. 
For all BHs we adopt an averaged spectrum obtained from observations of 108,104 low-$z$ QSOs \citep{Krawczyk2013}, which is a broken power law with a spectral index of 1 for $h_{\rm P}\nu > 200$ eV. We rescale it with the luminosity of each BH, which rarely exceeds the Eddington value. A notable exception is our brightest BH, found at $z=7$, which has an Eddington ratio of $2.9$. In general, as can be seen from Fig.~\ref{fig:SEDs}, the volume average of the BH spectra is brighter than that of other source types.
At $z=6$ we have a total of $2,745$ BHs. 
In Fig.~\ref{fig:LF} the BH LF at $z=6$ is shown, along with the observational data (and a fitting LF) from \cite{Giallongo2015}.
Although we are constrained by the box size and therefore do not sample the rarest, brightest BHs, we do however have a population of faint BHs. Our LF matches the range of brighter magnitudes where observational constraints are available.
In a companion paper, by means of a neural network applied to the BH population of the MBII simulation, we re-estimate the faint end of the LF (Eide et al., in prep).

\textit{X-ray Binary Systems (XRBs)}: As with the black holes, we separately obtain a spectrum and rescale it with the individual luminosities of the sources.
We look up spectra (evolving only with redshift) and luminosities (scaling with the physical properties of the sources) from the libraries of \cite{Fragos2013,Fragos2013a}, using the updated version presented by \cite{Madau2017}.  In particular, the luminosity scales with properties such as total mass, metallicity and age of the stellar particles, as well as the star formation rate in the halo.
Our models include contributions from high-mass XRBs, which trace star formation and are dominant at $z\gtrsim 2.5$, as well as low-mass XRBs, which trace the stellar mass and dominate at lower $z$.
Their spectra are generally hard, peaking at keV-scales, as shown in Fig.~\ref{fig:SEDs}. As their AB magnitude is negligibly small, we calculate their $z=6$ LF at a wavelength of $6.2$~\AA, corresponding to photon energies of $2$~keV, and include it in Fig.~\ref{fig:LF}. The peak of the $2$~keV LF of the XRBs is at a magnitude $\rm M_{2\,\rm keV} \sim 1$. 

\begin{figure}
\centering
\includegraphics[width=\columnwidth]{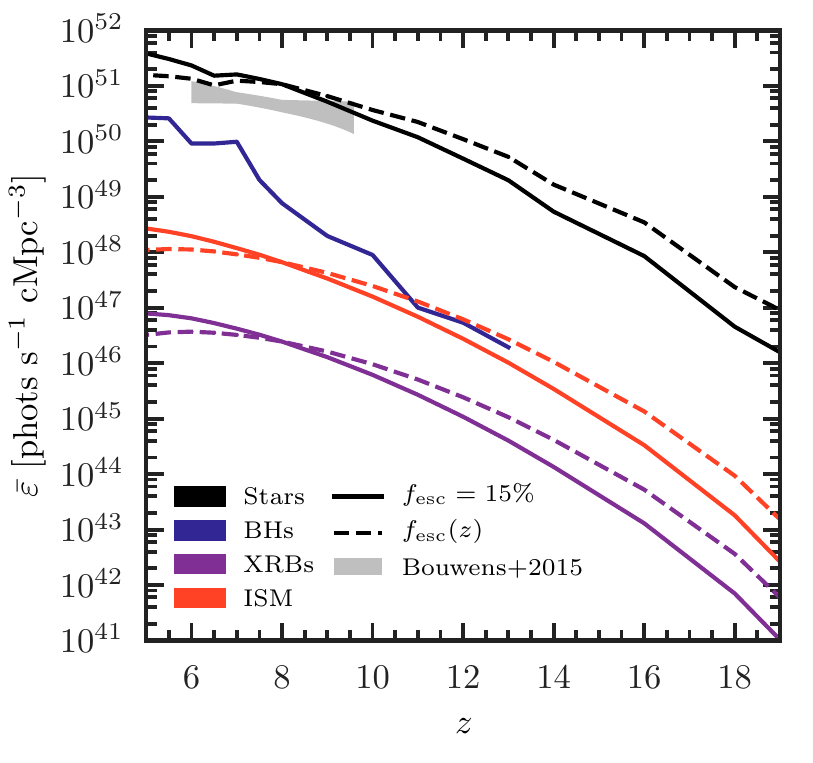}
\caption{Redshift evolution of the comoving volume averaged ionizing emissivity per source type
assuming either a constant UV escape fraction $f_{\rm esc} = 15\%$ (solid lines) or redshift-dependent $f_{\rm esc}(z)$ (dashed lines). From top to bottom the sets of curves correspond to: stars (black), BHs (blue), XRBs (purple) and ISM (red). The grey area corresponds to observationally constrained 95\% CI for the evolution of the cosmic ionizing emissivity~\citep{Bouwens2015ems}.}
\label{fig:emissivity}
\end{figure}

\textit{Diffuse thermal bremsstrahlung from shock heated ISM (ISM)}: Whereas XRBs are point sources of X-rays in galaxies, a diffuse X-ray component has also been observed \citep{Mineo2012a}. 
We model this as a redshift independent broken power-law spectrum which is flat until the characteristic thermal energy break at $k_{\rm B} T^{\rm ISM} = 240$ eV (observationally determined by \citealt{Mineo2012a}, where $k_{\rm B}$ is the Boltzmann's constant), and has a spectral index of 3 at higher energies. 
As the shock heating process is likely to be associated with supernovae \citep{Meiksin2017}, we scale the ISM luminosities of the galaxies with their star formation rate following \cite{Mineo2012a}. 
As seen in Fig.~\ref{fig:SEDs}, the spectra are fainter than the stellar ones below \HeII ionizing energies, while they are considerably softer than those of XRBs. From Fig.~\ref{fig:LF} we note that the ISM is much fainter than stars and BHs, as the brightest interstellar gas corresponds to the faintest BHs at ${\rm M}_{\rm AB} \sim -13$.

As we combine scaling relations from the literature with the physical properties obtained from MBII, the only real free parameter in our simulations is the escape fraction of ionizing photons.
While in Eide2018 we rescaled the emission of $h_{\rm P}\nu < 200$ eV photons by the constant factor $f_{\rm esc} = 15\%$, observational and theoretical investigations suggest that the averaged escape fraction of ionizing photons may have evolved with redshift along with the changing occurrence of environments that allow for their escape\footnote{See e.g. observations from \cite{Izotov2018}, \cite{Vanzella2018} and \cite{Fletcher2019}, in addition to suggestions of higher $f_{\rm esc}$ due to turbulence by e.g. \cite{Safarzadeh2016}, \cite{Kakiichi2019} and \cite{Kimm2019}, or of an environment-dependent $f_{\rm esc}$ as seen in simulations from e.g. \cite{Paardekooper2015}, \cite{Trebitsch2017} and \cite{Katz2018UVB}. A strong dependence is expected also on the evolution of the chemical properties of galaxies and in particular on the presence of dust (see e.g. \citealt{Chisholm_2018,He.Ricotti.Geen_2020,Graziani2020}).}.

In this work, we examine the effect of varying the escape fraction following the models and Bayesian constraints from Planck presented by \cite{Price2016},
\begin{equation}
f_{\rm esc}(z) = f_{{\rm esc},z=8} \left( \frac{1+z}{9}\right)^\beta,
\end{equation}
where $f_{{\rm esc},z=8}$ is the escape fraction at $z=8$.
We choose $f_{{\rm esc},z=8} = 15\%$ and $\beta = 2.2$, where the latter is slightly lower than the \cite{Price2016} lower limit. 
Our evolving escape fraction is essentially  $100\%$ at high-$z$ before decreasing to single-digit values at $z<8$.  

\begin{figure}
\centering
\includegraphics[width=0.97\columnwidth]{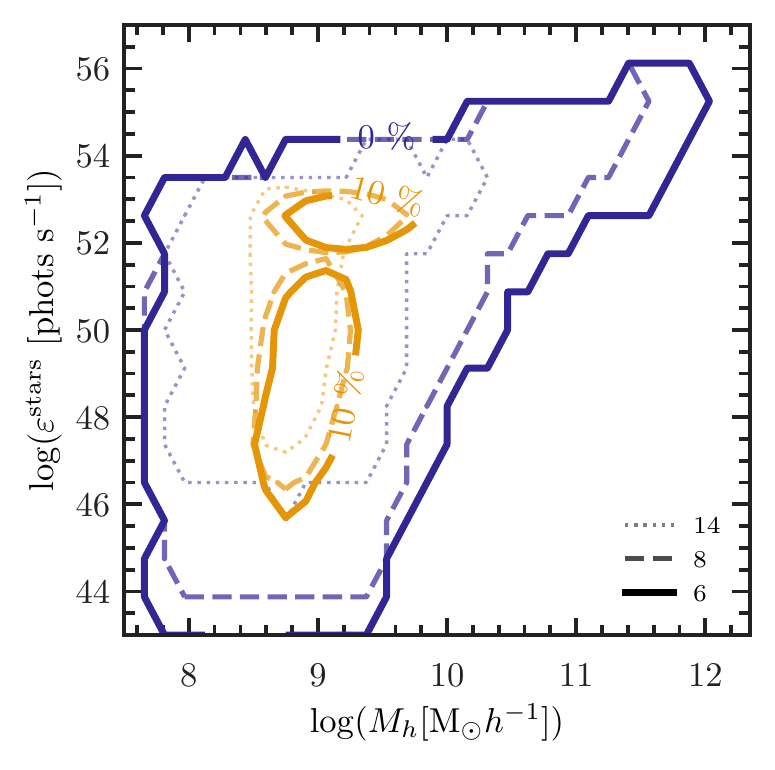}
\caption{Ionizing emissivities of stars, $\varepsilon^{\rm stars}$, as a function of their halo masses, $M_h$, at $z=14$ (thin dotted line), $8$ (dashed line) and $6$ (thick solid line). The contours are drawn at the 0th percentile (i.e. full distribution, blue lines) and 10th percentile (yellow lines).
}
\label{fig:Mhalo_dniondt}
\end{figure}

Our choices for the value of the escape fractions have also been guided to yield emissivities consistent with the \cite{Bouwens2015ems} measurements, as can be seen in Fig.~\ref{fig:emissivity}, where we show the evolution of the volume averaged ionizing emissivity  from the various source types. 
The difference between a constant and varying escape fraction is now reflected on the emissivity, which is higher at $z>8$ with $f_{\rm esc}(z)$.
Stars dominate the budget at all $z$, as expected, independently from the adopted escape fraction.
There is however an increasingly significant contribution from black holes at $z<10$, as their volume averaged emissivity increases by three orders of magnitude from the appearance of the first BH at $z=13$, and $z=6$. The contribution from the ISM is similar to that of BHs between $z=13$ and $11$ assuming $f_{\rm esc}(z)$, otherwise it is sub-dominant. The volume averaged emissivity of the XRBs is fainter by yet another order of magnitude compared to that of the ISM.

To investigate which halos contribute most to the ionization budget, in Fig.~\ref{fig:Mhalo_dniondt} we plot the stellar emissivity (i.e. produced by all stellar particles contained within a halo), $\varepsilon^{\rm stars}$, as a function of the hosting halo mass, $M_h$. Halos are identified with a friends-of-friends procedure and have a minimum mass of $\sim 9 \times 10^6 \, h^{-1}$~M$_\odot$. However, we have stars that have been associated (potentially unreliably) with even lower mass halos. We do not consider this to have significant impact on the reliability on the emissivity of the stars, as their properties are derived mainly from the star particles and not the halos. As $\varepsilon^{\rm stars}$ is assigned according to a number of physical properties, it has a strong degeneracy with $M_h$. The majority of the sources ($90\%$) are found in the narrow range of halo masses $8.5 < \log M_h/({\rm M}_\odot h^{-1}) < 9.6$ at all redshifts considered here, but their emissivities differ by eight order of magnitudes, indicating the strong contribution from e.g. stellar ages and metallicities to the stellar emissivity. The most massive halos, with $\log M_h/({\rm M}_\odot h^{-1}) > 11$, are few in number, but yield exclusively high emissivities. These results are also suggestive that adopting the same escape fraction for all halos is an oversimplification, a topic which will be investigated in more detail in a separate paper. 

Finally, we perform multifrequency 3D Monte Carlo radiative transfer simulations with the code \texttt{CRASH} \citep{Ciardi2001,Maselli2003,Maselli2009,Graziani2013,Graziani2018}.
We discretize the spectrum of each source into $82$ frequency bins between $13.6$ eV and $2$ keV, spaced closer around the ionization thresholds of hydrogen ($13.6$~eV) and helium ($24.6$ and $54.4$~eV), both of which \texttt{CRASH} tracks the ionization state of, in addition to the gas temperature\footnote{We use 72 bins below 200~eV and 10 in the range 200~eV--2~keV. We have verified that this choice assures convergent results.}.  
The emitted packets of photons are depleted and redshifted as they propagate  (see Eide2018 for a more detailed description), and they are assumed to be lost once they exit the volume. This assumption does not impact the results presented here (see Appendix~\ref{app:loss} for a more extensive discussion).

\section{Results}
\label{sec:results}

We first present some key qualitative findings that manifest themselves in all our results (\autoref{sec:qualitative_overview}), and then turn to present detailed reionization histories, phase diagrams and thermal properties of the IGM in our simulations with various combinations of different source types and escape fractions (\autoref{sec:reionization_history}). We finally study the results in light of various observational constraints (\autoref{sec:observational_constraints}).

\subsection{Qualitative overview}
\label{sec:qualitative_overview}

\begin{figure}
\centering
\includegraphics[width=\columnwidth]{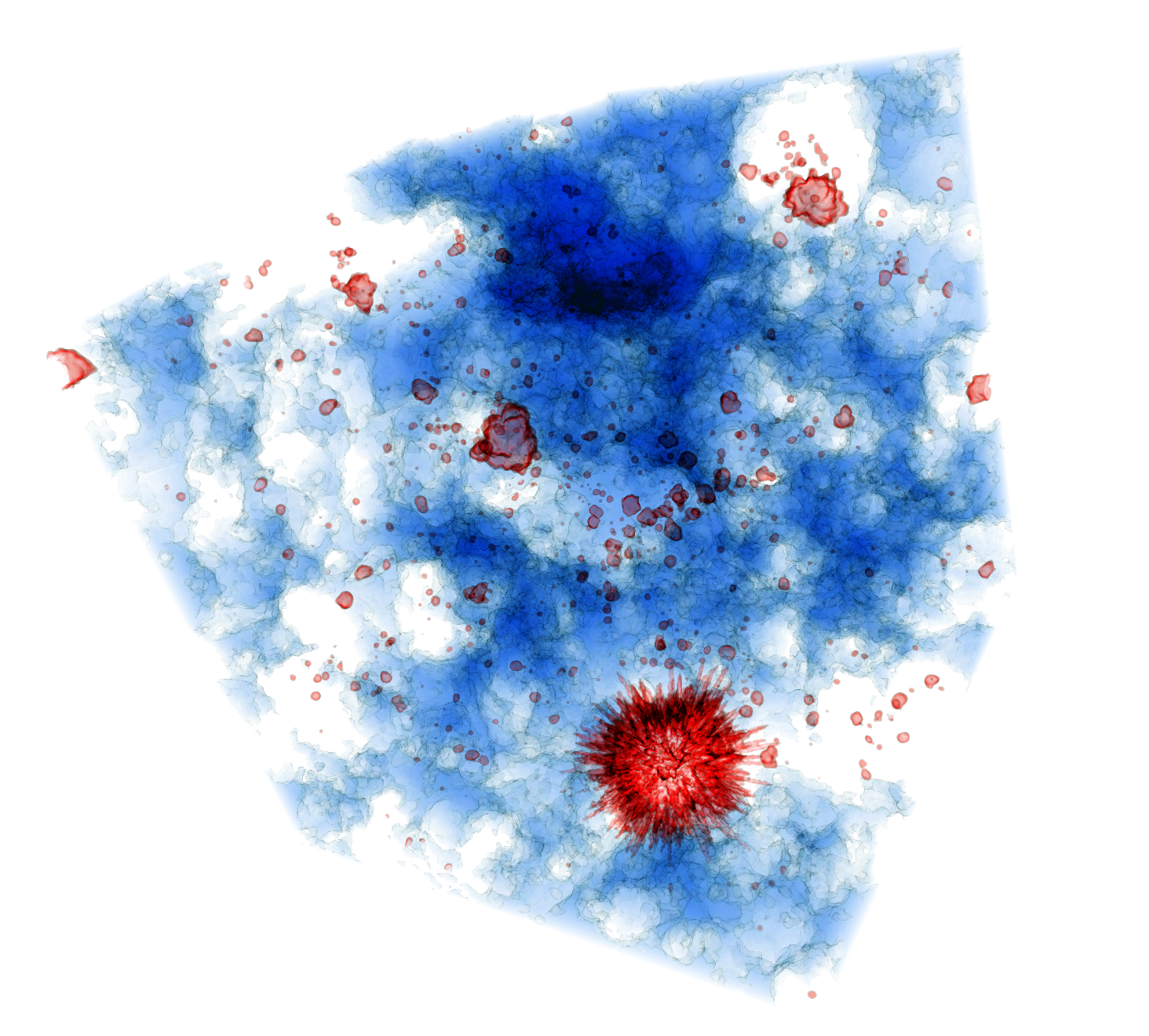}
\caption{Volumetric rendering at $z=6.9$ for a simulation with ionizing radiation from stars, BHs, XRBs and the ISM of galaxies, using a UV escape fraction $f_{\rm esc} = 15\%$. Each side of the volume is $100 h^{-1}$ cMpc long. The blue regions show the distribution of the neutral \HI and are drawn at $x_\HI > 0.1$, whereas red contours indicate $\xHeIII > 0.1$, which we only find around BHs. The brightest BH is located in the lower right corner.}
\label{fig:3D}
\end{figure}

In Fig.~\ref{fig:3D} we present a 3D rendering of the full ($100 h^{-1}$ cMpc)$^3$ volume at $z=6.9$, when the hydrogen reionization process nears completion and with helium reionization well underway\footnote{Note that the redshift displayed here is lower than those analysed in subsequent figures as in 3D more details can be discerned when the ionized fractions are higher. The presence of an ionized region at the location of a bright BH does not necessarily mean that the BH is active at all redshifts, but rather that the ionized region hasn't recombined yet.}. It shows the \HI and \HeII ionization state of the IGM for a simulation including all the source types and with $f_{\rm esc} = 15\%$. The volume averaged ionization fraction is $\vHII = 0.85365$ for hydrogen, and $\vHeIII = 0.00731$ for \HeII\footnote{We reach a convergence at the 10$^{-5}$ level. See \cite{Eide2018} for more details.}
The neutral IGM is perforated by ionized \HII regions, not necessarily centered on the brightest sources. Instead, their extent illustrates the effect of the ionization provided by stars. Multiple galaxies contribute jointly to enlarge the \HII fronts after individual \HII bubbles merge. The morphology thus displays signatures of the percolation inherent to the \HI reionization process \citep{Furlanetto2016,Bag2019}, a trait which will be investigated in more detail in a companion paper (Busch et al.~sub). The \HII regions host \HeIII bubbles of various extent, most centered on BHs, indicating their responsibility in driving \HeII reionization. The figure highlights the effect of our most luminous BH, which distinguishes itself through a large \HeIII region. The spikes associated to it are mainly due to recently ionized channels of lower optical \HeII depth, but there is also a smaller component associated to numerical artefacts of a yet unconverged, non-equilibrium physical state (see also the physical properties of the environment of the high-$z$ BH modelled by \citealt{Kakiichi2017}).

\begin{figure*}
\includegraphics[width=0.82\textwidth]{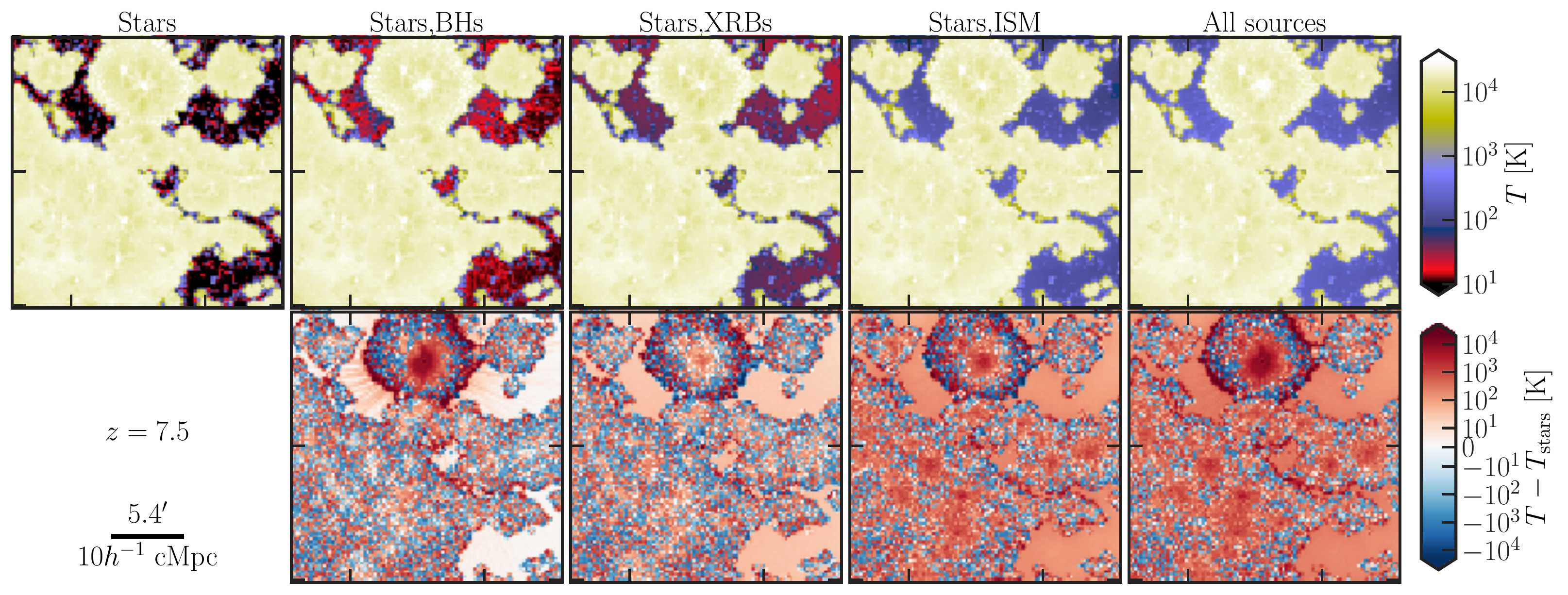} 
\includegraphics[width=0.82\textwidth]{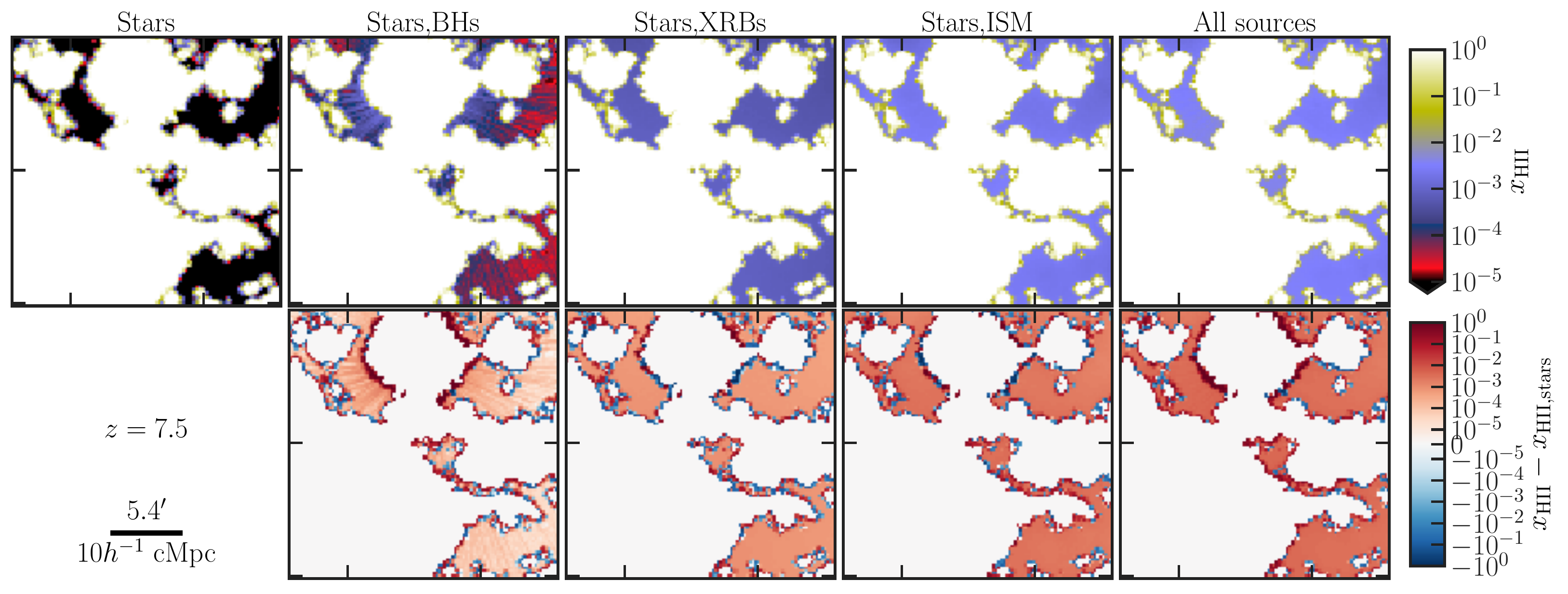} 
\includegraphics[width=0.82\textwidth]{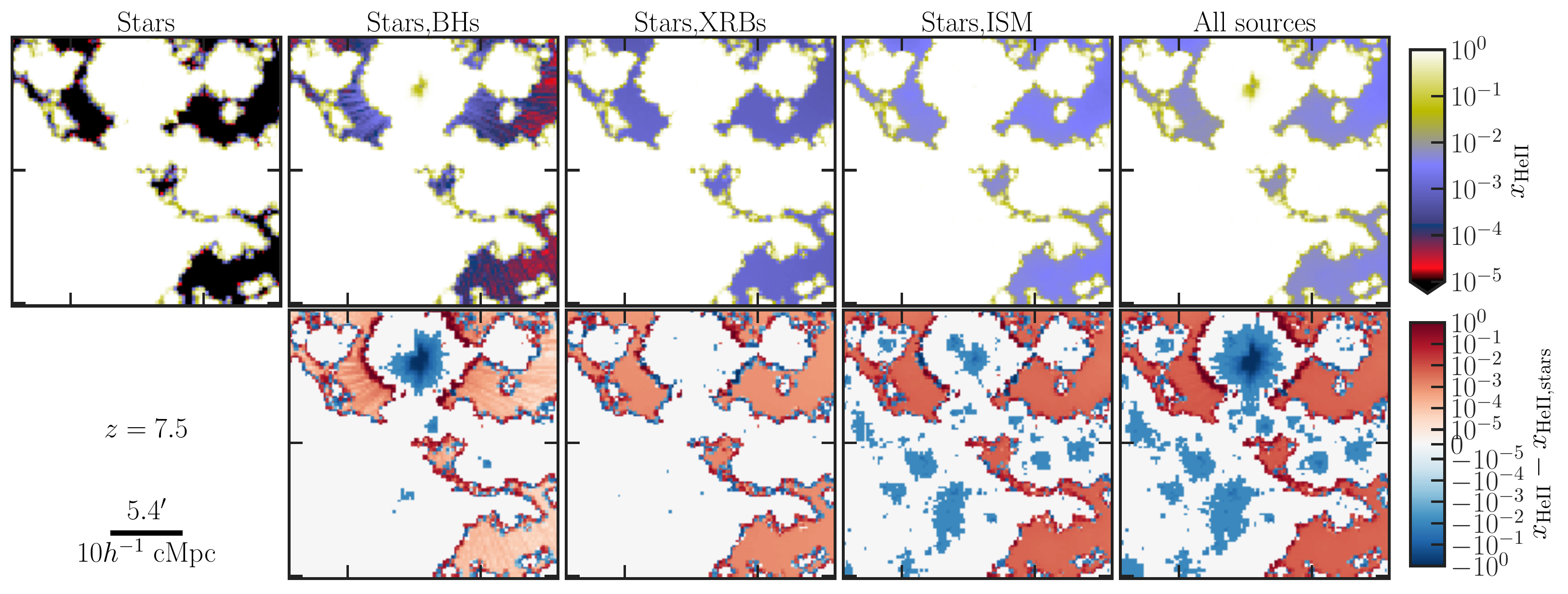} 
\includegraphics[width=0.82\textwidth]{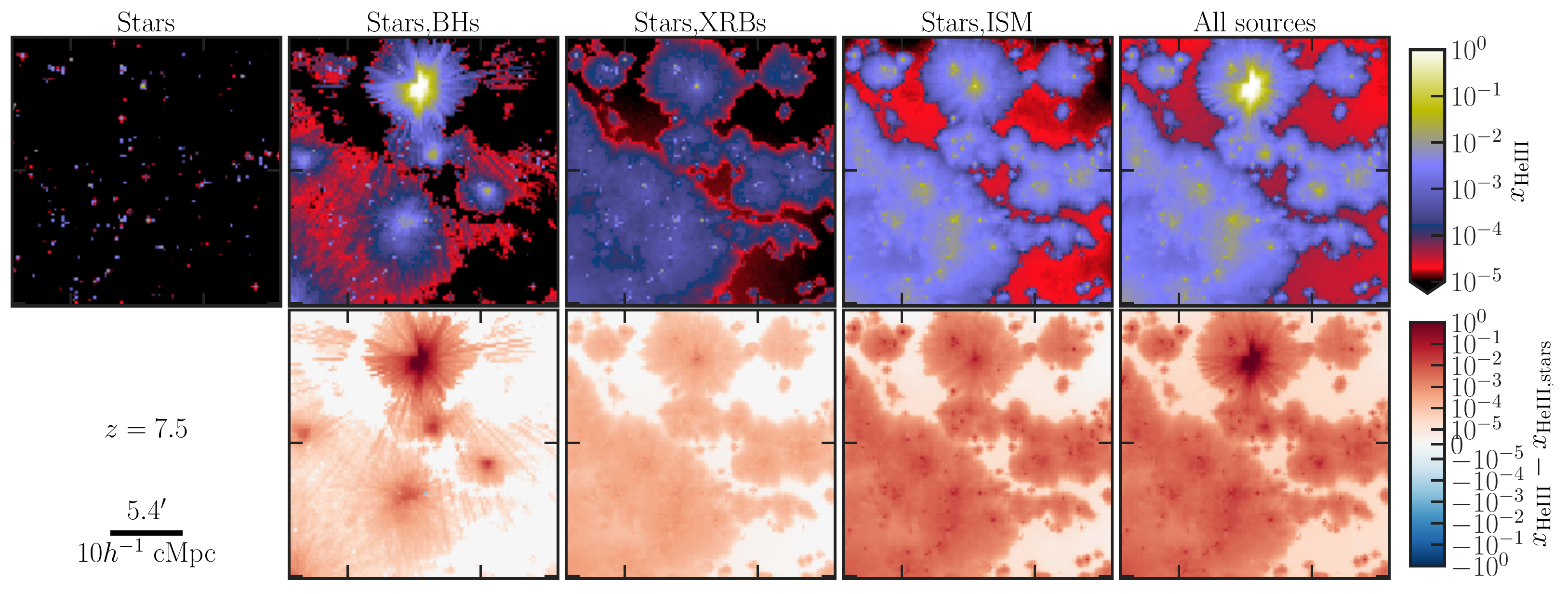} 
\caption{From top to bottom maps of the temperature, \HII, \HeII and \HeIII fractions of the gas surrounding the location of a BH at $z=7.5$ for different combinations of ionizing sources (left to right): stars, stars and BHs, stars and XRBs, stars and ISM, all sources. The lower sets of rows show the difference with respect to the stars alone simulations. The maps are $25 h^{-1}$ cMpc wide. Note that alternating blue/red pixels at the edges of HII regions (and within the ionized regions in the temperature maps) are due to Monte Carlo noise and that only the red ring of extra ionization/heating produced by the bright quasar has a physical meaning. See text for more details. 
}
\label{fig:maps}
\end{figure*}

In Fig.~\ref{fig:maps} we show maps of the thermal and ionization state of the IGM at $z=7.5$, centered around the brightest BH seen in Fig.~\ref{fig:3D}. The BH resides in a $5.76 \times 10^{11} {\rm M}_\odot$ halo, has an AB magnitude at $1450$ Å ${\rm M}_{\rm AB} = -21.9$, a mass $M_{\rm BH} = 1.85 \times 10^{7} {\rm M}_{\odot}$ and an ionizing emissivity $\varepsilon = 1.13 \times 10^{55}$ phots s$^{-1}$. This is not as bright or massive as ULAS J1120+0641 at $z=7.54$ (which has ${\rm M_{\rm AB}} = -26.76 \pm 0.04$ and a mass $M_{\rm BH} = 7.8^{+3.3}_{-1.9} \times 10^8 {\rm M}_\odot$; \citealt{Banados2018}), but it nevertheless belongs to the brightest end of the BH LF that can be sampled with a $100 h^{-1}$ cMpc volume.
The maps illustrate a few important concepts that will be further discussed in the following sections. The extent and temperature of $\sim 10^4$~K of the \HII regions are determined by the stars. Additional and significant \HII heating is only provided when \HeII gets fully ionized by the BHs (see also \citealt{Kakiichi2017} for further discussion), although
partial ionization and heating around fully ionized regions is also provided by the shock heated ISM and, to a lesser extent because of the harder spectrum, the XRBs (see also \citealt{Graziani2018}). This will have a strong impact on 21~cm \HI observations, which will be examined in a companion paper (Ma et al.~in prep). Additional heating is also clearly visible within \HII regions, in correspondence to a substantial presence of \HeIII, i.e. in the immediate surroundings of the more energetic sources. The alternating blue/red pixels correspond to Monte Carlo noise, which is particularly evident in the temperature maps within the \HII regions. This happens because it is statistically impossible for two cells with $\xHII=1$ to have the same exact temperature in simulations with different source types.

\subsection{Reionization and reheating history}
\label{sec:reionization_history}
\begin{table*}
  \begin{threeparttable}
    \caption{Thermal and ionization state of the IGM at $z=6$ for different combination of source types and escape fractions. From left to right, the columns refer to the volume averaged fraction of \HII, \HeII, \HeIII, \HI, the volume averaged temperature, $\vT$, and the temperature at mean density, $T_0$. The first and second value in each column refers to $f_{\rm esc} = 15$\% and $f_{\rm esc}(z)$, respectively.
    }
\label{tab:IGM_state}
     \centering
     \begin{tabular}{cllllll}
        \toprule
        Source type & $\vHII$ & $\vHeII$ & $\vHeIII$ & $\vHI$ & $\vT$ (K) & log($T_0$)/K\\
        \midrule
    Stars & 0.99998/0.99852 & 0.99971/0.99829 & 0.00019/0.00016 & 0.00002/0.00148 & 19,274/18,643 & 4.269/4.256\\ 
    Stars, XRBs & 0.99998/0.99854 & 0.99903/0.99762 & 0.00087/0.00085 & 0.00002/0.00146 & 19,374/18,746 & 4.271/4.258\\
    Stars, ISM & 0.99998/0.99869 & 0.98982/0.98889 & 0.01008/0.00972 & 0.00002/0.00131 & 19,975/19,350 & 4.284/4.272\\
    Stars, BHs & 0.99998/0.99923 & 0.98174/0.98110 & 0.01816/0.01804 & 0.00002/0.00077 & 19,524/18,915 & 4.275/4.263\\
    All sources & 0.99998/0.99935 & 0.97050/0.97035 & 0.02940/0.02892 & 0.00002/0.00065 & 20,351/19,751 & 4.293/4.281\\   
\bottomrule
\end{tabular}
\end{threeparttable}
\end{table*}

\begin{figure}
\centering
\includegraphics[width=0.91\columnwidth]{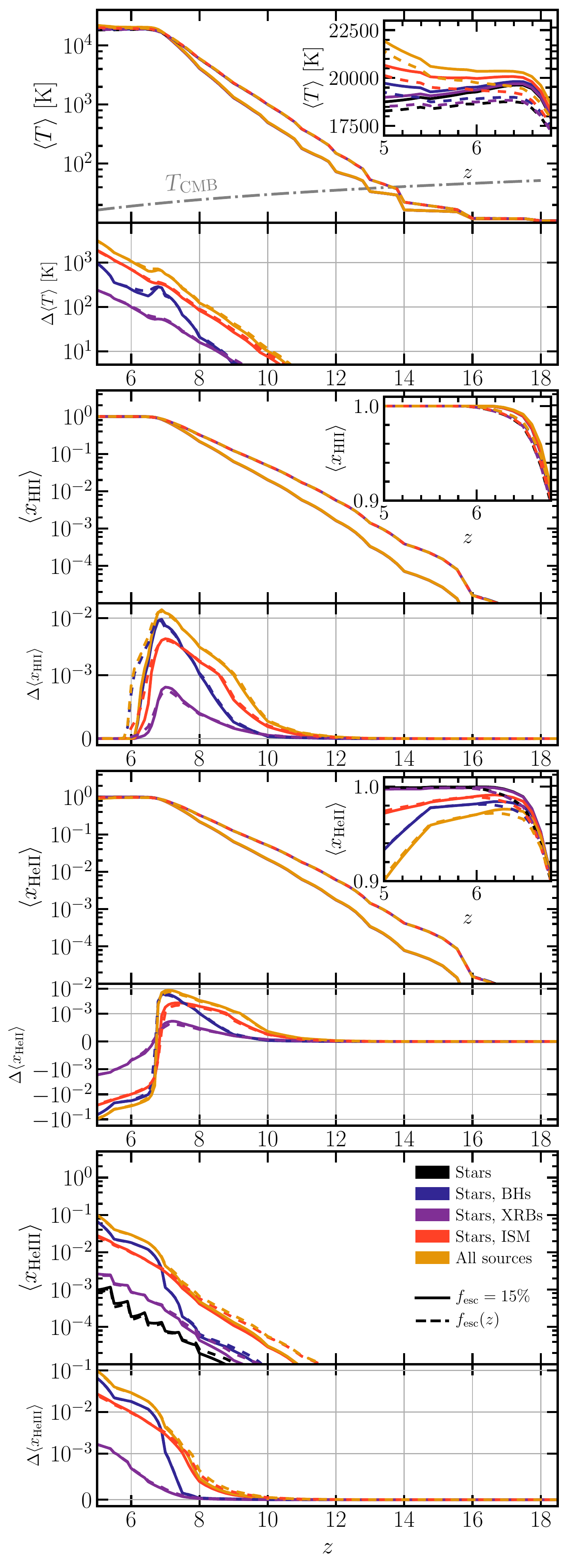}
\caption{Evolution of the volume averaged (from top to bottom) temperature, fraction of \HII, \HeII and \HeIII for a constant $f_{\rm esc} = 15$\% (solid lines) and an evolving $f_{\rm esc}(z)$ (dashed). The line colour refers to the ionizing sources included: stars (black), stars and BHs (blue), stars and XRBs (purple), stars and ISM (red), and all sources (yellow). The lower panels show the differences with respect to simulations with only stars. The dotted-dashed grey line in the top panel is the CMB temperature.}
\label{fig:reion_history}
\end{figure}

In Fig.~\ref{fig:reion_history} we present the reionization and thermal histories  for our five combinations of source types. 
The onset of \HI and \HeI reionization occurs at $z \sim 16$ for all models, and slightly earlier with an evolving escape fraction. 
With $f_{\rm esc}(z)$ the volume averaged ionization fractions remain higher until $z \sim 7$, when $f_{\rm esc}$ falls below $15\%$ and the trend is reversed. 

The main driver of the evolution of $\vHII$ and $\vHeII$ are the stars, with a non negligible contribution from the ISM and, at $z<8$, the BHs (see Fig.~\ref{fig:emissivity}). The ionization fractions differ most at $z \sim 7$, but they converge again to similar values towards the end of reionization, when $\vHII \sim 1$.
We also observe that the \HeII fraction below $z=7$ with stars only is higher than with more energetic sources, as these are starting to produce an appreciable amount of \HeIII, visibly depleting \HeII.  

The onset of \HeII reionization is strongly dependent on the spectral hardness of the sources, occurring between $z\sim 11$ (when ISM or all sources are included) and $z\sim 9$ (with stars only). While XRBs increase $\vHeIII$ by a factor of about five compared to stars only, including the contribution from the ISM or the BHs increases the difference to two order of magnitudes, the latter becoming the dominant source of \HeII reionization at $z<8$. 
With the exception of the initial and final stages of reionization, the evolution of the volume averaged temperature and ionization fractions are approximately exponential.

\begin{figure}
\centering
\includegraphics[width=0.9\columnwidth]{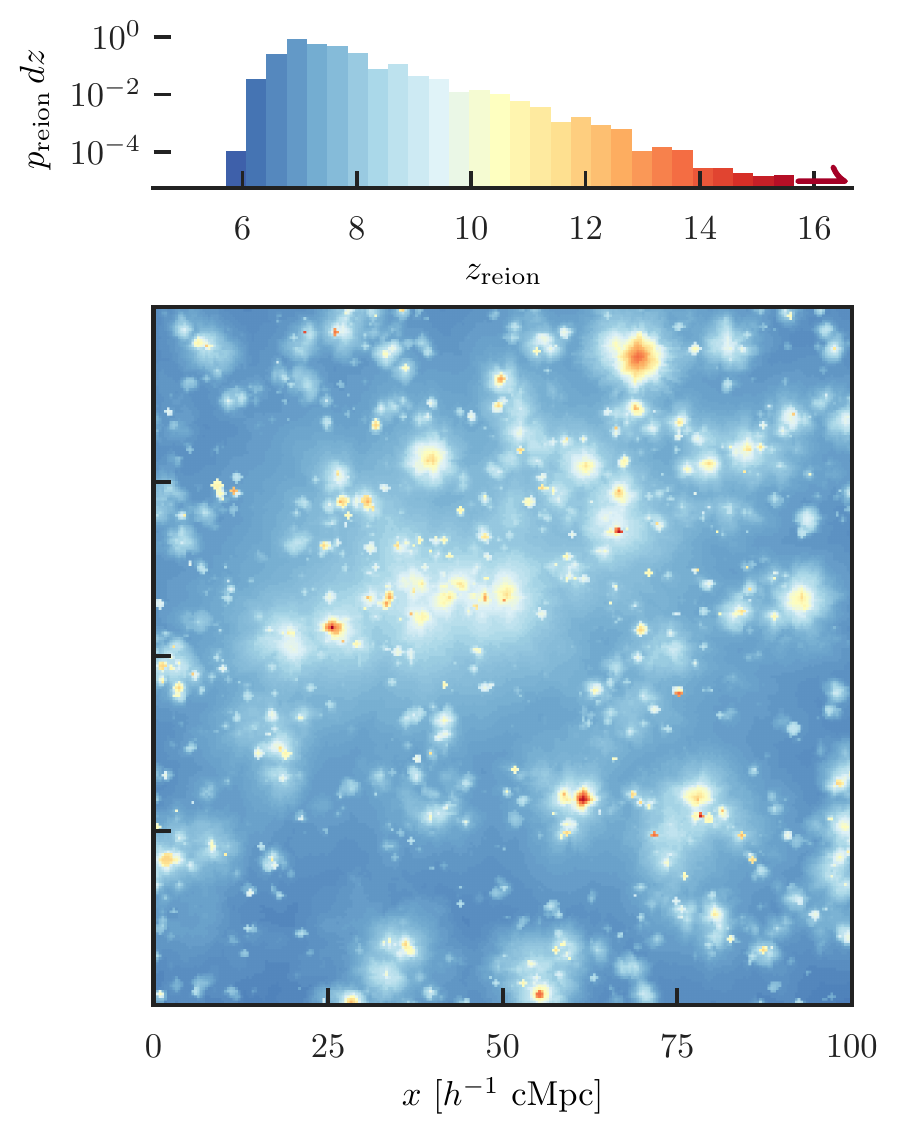}
\caption{Slice of the simulation with stars only and $f_{\rm esc} = 15\%$ showing the reionization redshifts $z_{\rm reion}$, colour coded according to the values in the histogram. The histogram indicates the distribution of $z_{\rm reion}$, given as the normalized probability density $p_{\rm reion}\, dz$.} 
\label{fig:z_reion}
\end{figure}

By $z\sim 6$ both \HI and \HeI reionization are concluded globally, with $\vHII \approx \vHeII \sim 1$ (see Table~\ref{tab:IGM_state}). However, the timing of reionization, $z_{\rm reion}$, varies with location, as it is clear from Fig.~\ref{fig:z_reion}, where $z_{\rm reion}$ (defined as the redshift at which $\xHII$ becomes larger than 0.9) is shown through a slice of our simulation with only stellar sources and $f_{\rm esc} = 15\%$ (see Fig.~\ref{fig:delta_z_reion} for a comparison with the other simulations). Reionization occurs earlier closer to the sources, supporting an inside-out scenario. 
Hydrogen in a small fraction of the IGM gets completely ionized at $z>12$, including the one surrounding the massive BH discussed earlier, but reionization occurs at $z \sim 7$ for the majority of the IGM. Lowering the ionization threshold to $\xHII \geq 0.1$ does not alter $z_{\rm reion}$ for most of the IGM, except that which is closest to the sources, where $z_{\rm reion}$ could increase from $z\sim 10$ to $z\sim 12$, as detailed in App.~\ref{app:reionization_timing}.
We defer a more thorough discussion of the correlation between the timing of reionization and its sources to a separate paper.

Turning to the reheating history, its global onset, evolution and conclusion follows closely that of \HI/\HeI reionization. The temperature of the \HII/\HeII regions is determined mainly by stellar type sources, while the harder sources heat the IGM surrounding the fully ionized regions (as seen from the maps in Fig.~\ref{fig:maps}). The volume averaged impact of these energetic sources is minor, but nevertheless visible, as the temperature is systematically higher in their presence (see Table~\ref{tab:IGM_state}). Similarly to what we found for \HeIII, the difference with the stellar case only increases with decreasing redshift.
Consistently with the trend observed in the ionization fractions, the temperatures are always higher with an evolving escape fraction. 
From the inset in Fig.~\ref{fig:reion_history}, at $z<6.5$ we observe a clear decline in temperature for all the simulations with $f_{\rm esc} = 15\%$, except when all sources are included, in which case it rises from $\vT = 20,277$~K at $z=6.5$, whereas this cooling is less evident with $f_{\rm esc}(z)$. The cooling is indicative of the transition towards an expected thermal asymptote \citep{Hui1997} which can be upset and postponed by \HeII reionization, during which further heating is provided \citep{McQuinn2009}. At $z<6$ we see the clear thermal signature of the latter in all our simulations except for those with stars alone, or stars and XRBs, both of which are inefficient \HeII ionizers.

\begin{figure*}
\centering
\includegraphics[width=\textwidth]{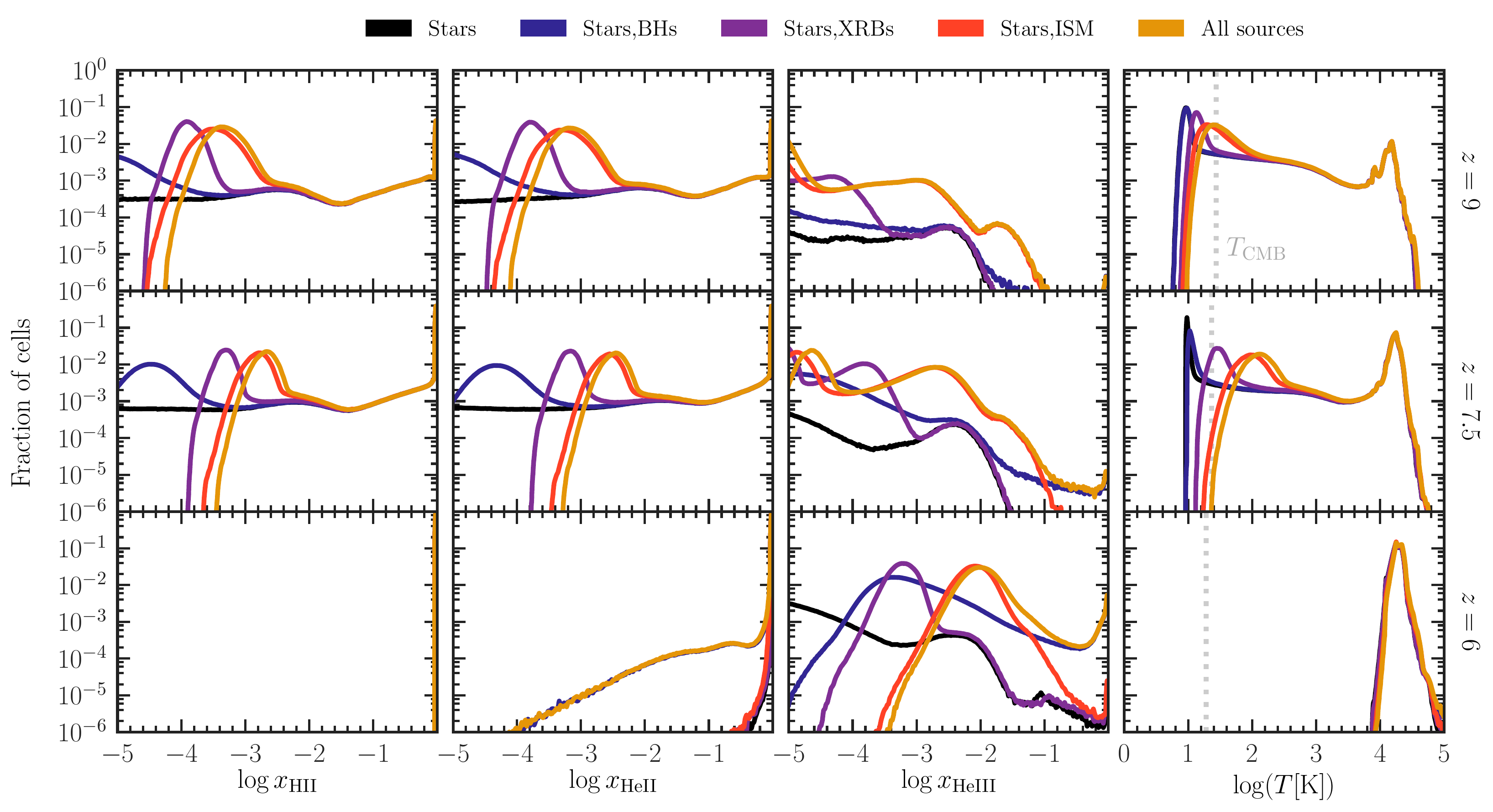}
\caption{Distributions (from left to right) of the \HII, \HeII, \HeIII fractions, and the temperature at (from top to bottom) $z=9$, $7.5$ and $6$. The line colour indicates the type of sources considered: stars (black), stars and BHs (blue), stars and XRBs (purple), stars and ISM (red), and all source types (yellow). The vertical dashed grey line in the temperature panels indicate the CMB temperature, $T_{\rm CMB}$.
}
\label{fig:pdf_quantities}
\end{figure*}

In Fig.~\ref{fig:pdf_quantities} we plot the distributions of the various quantities characterising the physical state of the IGM at three different redshifts through reionization, $z=9$, $7.5$ and $6$.
We find pronounced bimodalities for all quantities except \HeIII and while the IGM still has neutral hydrogen and helium.
As previously discussed, full ionization of hydrogen and singly ionized helium, and heating of the gas to $T \gtrsim 10^4$~K, is caused by stellar type sources.
The IGM that is not ionized, is coldest and most neutral with stellar type sources. Of the other source types in addition to stars, we find more ionization of \HI and \HeI and heating with BHs, which is further increased by the presence of XRBs. The most partial ionization and heating, though, is provided by the ISM. 
The distribution of \HeIII is complex. With stars, $\log \xHeIII < -1.4$ at $z=9$ and $7.5$, and higher ionization is barely seen at $z=6$. The BHs do not provide large amounts of \HeIII until $z=6$, when they are responsible for the vast majority of fully ionized \HeIII. The XRBs produce \HeIII in appreciable quantities, but at lower $\xHeIII$ values than the stars, while the ISM provides copious amounts of \HeII ionization resulting in two peaks which shift towards higher values with decreasing redshift, but very little fully ionized helium.

We find gas with temperatures and ionization states in between the two peaks in the \HII, \HeII and $T$ distributions. This gas may be a numerical artefact, as it is unlikely that such gas exists with stellar sources, as discussed by e.g.~\citet{Ross2017} and \citet{Eide2018}.
Whenever the cell size is too large to resolve the sharp ionization front expected from stellar type sources, the cell containing the front appears partially ionized and warm, as in our simulations, while in reality part of the gas in the cell should be neutral and cold, and part fully ionized and hot. 
We discuss this issue more in detail in \citet{Ma2020} and Ma et al. (2020b, in prep).

\begin{figure*}
\centering
\includegraphics[width=\textwidth]{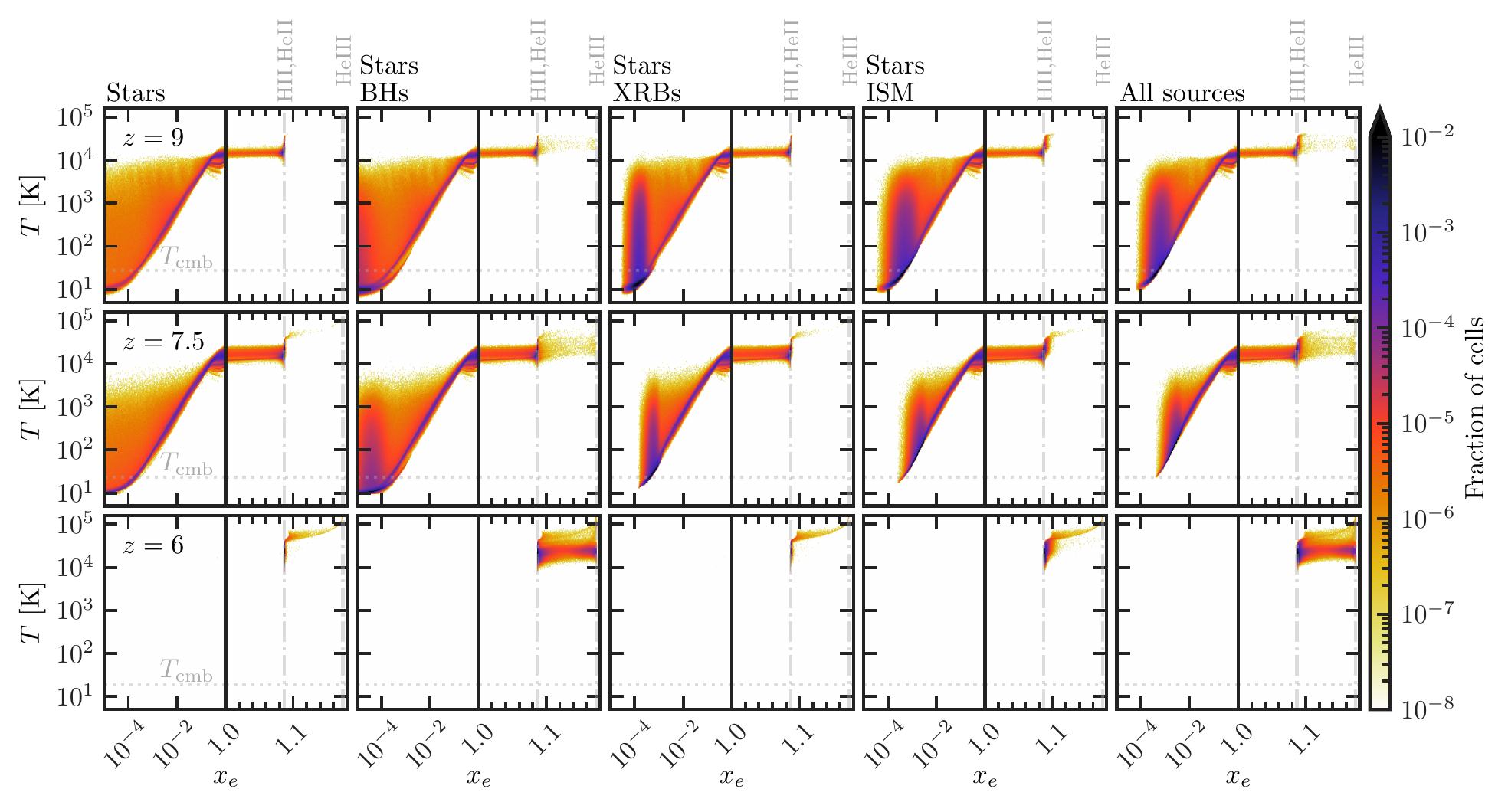}
\caption{Distribution of the temperature, $T$,
and the free electron fraction, $x_e$, for simulations with different combination of sources and $f_{\rm esc} = 15\%$ at $z=9$ (upper row), $z=7.5$ (middle row) and $z=6$ (lower row). From left to right, the panels refer to stars only, stars and BHs, stars and XRBs, stars and the ISM, all sources together. The dotted horizontal lines indicate the CMB temperature, $T_{\rm CMB}$, while the dash-dotted vertical lines refer to full \HI and \HeI ionization ($x_e = 1.09$), and to full \HI and \HeIII ionization ($x_e = 1.17$), respectively.
}
\label{fig:thermal_state}
\end{figure*}

In Fig.~\ref{fig:thermal_state} we show the distribution of the volume with a given temperature $T$ and free electron fraction
\begin{equation}
    x_e \equiv \frac{n_e}{n_{\rm H}} = \xHII + \frac{Y}{X} \left( \xHeII + 2 \xHeIII \right),
    \label{eq:free_e_frac}
\end{equation}
where $n_e$ and $n_{\rm H}$ are the number densities of free electrons and of hydrogen, while $X$ and $Y$ are the hydrogen and helium number fractions, respectively. 
As a reference, $x_e = 1.00$ when H is fully ionized and He is fully neutral, $x_e = 1.09$ when H is fully ionized and He is singly ionized, and $x_e = 1.17$ when both H and He are fully ionized. 
We plot the relations at three different redshifts, $z=9$, $7.5$ and $6$, showing distributions in which the gas is at various stages in transforming from cold ($T\sim 10$~K) and largely neutral ($x_e < 10^{-4}$) to hot ($T\sim 10^4$~K) and ionized ($x_e > 1$). At $z=9$, all the simulations have a significant amount of gas with $x_e < 10^{-2}$ and temperatures ranging from $10$~K to $\sim 10^3$~K, but still with significant amounts below $T_{\rm CMB}$. At $z=7.5$, only the simulations with either stars alone or stars and BHs have gas with $T<T_{\rm CMB}$. At $z=6$, all the gas in all the simulations has transitioned into an ionized state.

The highest temperatures (above $\sim 10^{4.5}$~K) are reached in regions where both H and He are ionized.  Their abundance depends on the spectral shape of the ionizing photons and it is maximum in the presence of BHs. In fact, only BHs are able to fully ionize He (see third column of Fig.~\ref{fig:pdf_quantities}), although as a whole their contribution to ionization and heating is marginal because of their paucity. A slight amount of gas in this physical state is also present with stars only at $z=7.5$ and $6$ (see also the upper panel of Fig.~\ref{fig:maps}), and corresponds to cells hosting sources\footnote{A negligible fraction of these cells are shock heated.}. 

The distribution of temperatures in the \HII/\HeII regions ($1 < x_e < 1.09$) is very similar in all simulations, ranging between $\sim$8,000~K and $\sim$20,000~K. The similarity confirms that stars are responsible for the physical state of the gas in such regions. The spread of temperatures observed in Fig.~\ref{fig:thermal_state} is indicative of the time elapsed since the ionization of the gas. In fact, under purely adiabatic expansion and assuming no recombination, $T(z) \propto (1+z)^2$. Furthermore, the cooling rate increases for temperature above $\sim 10^4$~K. The net effect is that gas that has recently been ionized is hotter than gas that has experienced ionization at an earlier time. This points to the gas temperature being a possible archaeological tracer of the reionization timing, supporting the work of e.g.~\cite{Keating2018}, who also found recently ionized regions to be hotter than earlier ionized ones.

\begin{figure}
    \centering
    \includegraphics[width=\columnwidth]{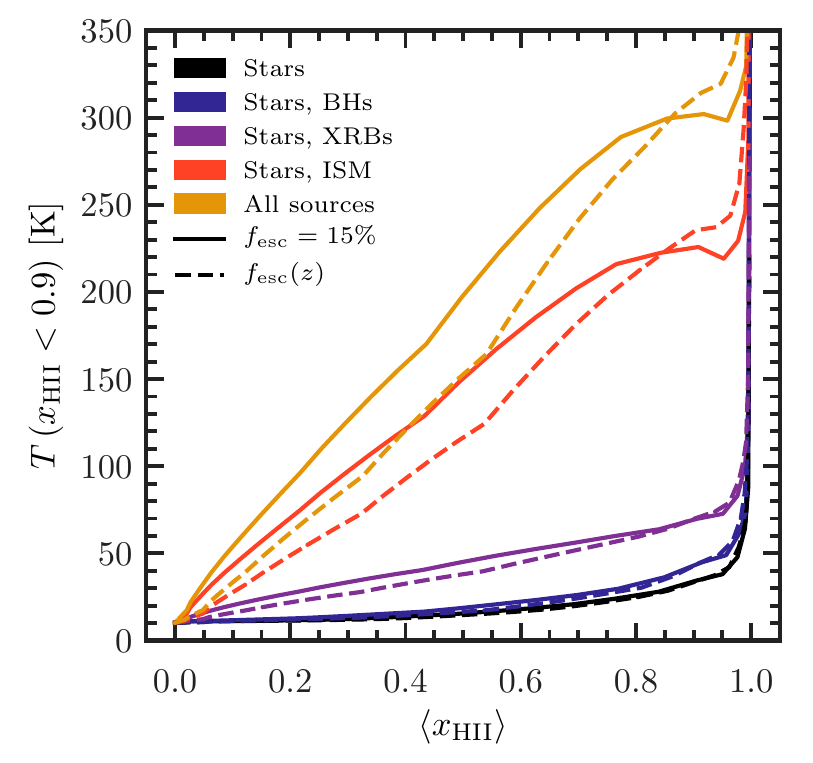}
    \caption{Median temperature of the gas with $\xHII \leq 0.9$ as a function of $\vHII$. The solid and dashed lines refer to simulations with a constant $f_{\rm esc} = 15\%$ and a varying $f_{\rm esc}$, respectively. The line colour gives the combination of source types (black: stars; blue: stars, BHs; purple: stars, XRBs; red: stars, ISM; yellow: all sources).}
    \label{fig:T_neutral}
\end{figure}

As the temperature of the neutral and partially ionized IGM is of great importance to the cosmological \HI $21$~cm signal, we will discuss this further in the following. 
In Fig.~\ref{fig:T_neutral} we plot the median temperature of gas with $\xHII < 0.9$ against the volume averaged $\vHII$ for our different source types and escape fractions.
The simulations reach $\vHII \sim 0.1$ 
at $z=8$--$10$, when the CMB temperature ranges between 24.5 and 30~K. 
As expected, little or no heating of this largely neutral gas is observed with stars only. In spite of their harder photons' ability to penetrate deeper into the IGM, the contribution from BHs is negligible (a few degrees more than the stars) due to their paucity. The more numerous XRBs and ISM, individually fainter than BHs (see Fig.~\ref{fig:LF}), are able to raise the temperature to $\sim 70$~K and $\sim 220$~K, respectively, as $\vHII$ approaches 0.9. An additional $\sim 50$~K are gained when all sources are taken together. With stars alone, the temperature of the neutral IGM will always be below $T_{\rm CMB}$, whereas with BHs it will be patchy with temperatures above in the vicinity of BHs and below further away from these sources. As for the XRBs and ISM, we find $T_{\rm gas} \gg T_{\rm CMB}$ at all \vHII values plotted in the figure, corresponding to redshifts $z \lesssim 10$.
This behaviour has important implications for the 21cm signal, which will be presented in Ma et al. (2020b, in prep). 

Simulations with $f_{\rm esc} = 15\%$ have temperatures higher than those with $f_{\rm esc}(z)$ at a fixed $\vHII$ as long as $\vHII \lesssim 0.85$, while at higher ionization fractions the trend is reversed, as $f_{\rm esc}(z)$ at $z<8$ reaches values lower than 15\%, resulting in a hardening of the spectra in simulations with a varying escape fraction compared to those with $f_{\rm esc} = 15\%$, i.e. higher temperatures in the neutral parts of the IGM. 

\subsection{Observational constraints}
\label{sec:observational_constraints}

In this section we will discuss our results in the context of available observational constraints.

\subsubsection{Abundance of neutral hydrogen}
\label{sec:HI}

\begin{figure}
\includegraphics[width=\columnwidth]{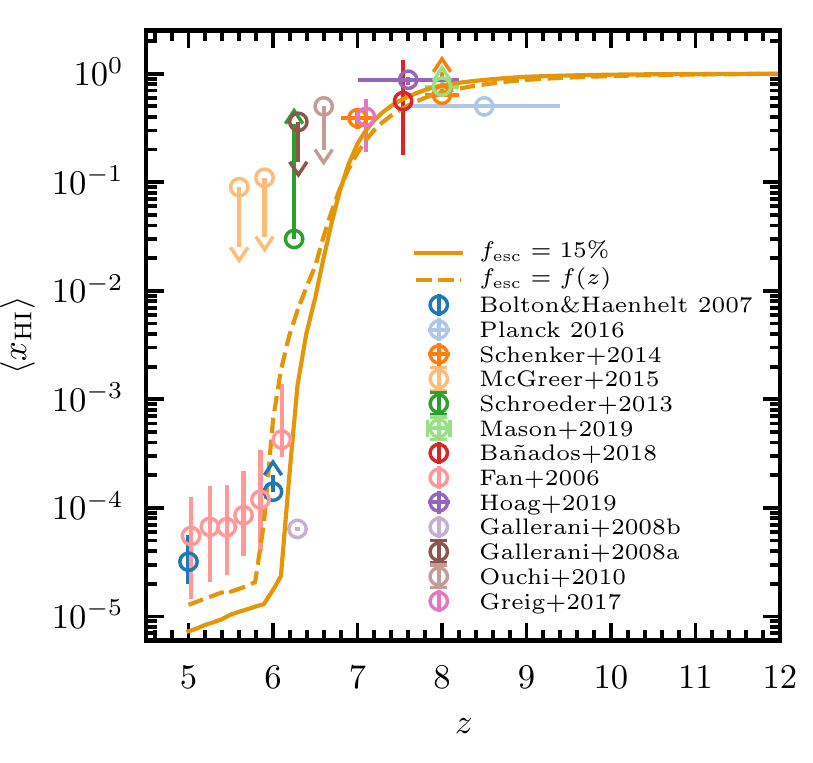}
\caption{Evolution of the volume averaged neutral hydrogen fraction, $\vHI$. The lines refer to simulations with all sources and either a constant $f_{\rm esc} = 15$\% (solid line) or an evolving escape fraction (dashed
line). Observational constraints are also shown. These are discussed in the main text.}
\label{fig:reion_conclusion}
\end{figure}

In Fig.~\ref{fig:reion_conclusion} we plot the evolution of $\vHI$ at $z<12$ for simulations including all source types and the two different escape fractions (for reference, in Table~\ref{tab:IGM_state} the values for all simulations are reported). Although the timing of full reionization is roughly the same in both simulations, there are some differences in the residual neutral fraction towards its conclusion, with $\vHI = 6.5 \times 10^{-4}$ ($1.8 \times 10^{-5}$)  at $z=6$ for $f_{\rm esc}(z)$ ($f_{\rm esc} = 15\%$). 
$\vHI$ remains approximately constant from $z=12$ until $z=9$, when it starts to decrease visibly. Consistently with what observed in Fig.~\ref{fig:reion_history}, $\vHI$ is lower for simulations with $f_{\rm esc}(z)$ until $z \sim 7$, and then the trend is reversed, with a difference at $z=6$ of about two orders of magnitude, as below $z=6.5$ $f_{\rm esc}(z)$ has reached the single-digit percent level.

Our results are consistent with observational data at $z \gtrsim 7$, reproducing the $\vHI$ derived from absorption lines of the $z=7.54$ QSO ULAS J1342+0928 by \cite{Banados2018}, the $\vHI$ statistically inferred from damping wing absorption in the spectra of ULAS J1120+0641 \citep{Greig2017}, the lower limit of \cite{Mason2019} from the lack of \Lya emission in 53 faint LBGs at $z>7.2$, the constraints from 68 LBG candidates of \cite{Hoag2019}, and also the knee at $z \sim 7$ matches well the neutral fraction of \cite{Schenker2014} inferred from observed LBG visibilities. 

At $z<7$ our results are well within the upper limits from the dark pixel analysis of \cite{McGreer2015}, from the LAE clustering by \cite{Ouchi2010}, as well as the QSO sample analysis of \cite{Gallerani2008QSO}. Nevertheless, at $z\sim 6$ we observe some tension. More specifically, the results with $f_{\rm esc} = 15$\% are marginally consistent with the value of  the neutral fraction derived along the line of sight to a GRB at $z=6.29$ by \cite{Gallerani2008GRB}, but both are at odds with the \cite{Fan2006} \HI trough constraints at $z \leq 6$ and the QSO near-zones explored by \cite{Bolton2007}, suggesting e.g. that toward the end of reionization a lower escape fraction would be more appropriate. The simulation with an evolving  $f_{\rm esc}$ in fact seems to match better the observational constraints at $z \lesssim 6$. 
We should note that our estimate of the neutral fraction at these redshifts is likely an upper limit, as we do not resolve nor model as sub-grid physics the effect of small-scale Lyman-limit systems (LLS), which are instead crucial in the final stages of reionization as well as in the post-reionization IGM, as discussed e.g. by \cite{Madau2017LLS}. The inclusion of LLS would also help to reconcile both our models with the QSO damping wing analysis by \cite{Schroeder2013}, which indicates that a substantial part of the IGM must still be neutral at $z=6$. Such neutral patches may also be required to explain the long troughs observed in front of QSOs \citep{Chardin2017}.  

Finally, our timing of the reionization midpoint, $z_{50\%}$, when the volume average of the free electron fraction reaches $\langle x_e \rangle = 0.5$, is $7.5$ for $f_{\rm esc}=15\%$ (independent of source types) and $7.7$ for $f_{\rm esc}(z)$, in both cases within the constraints of
\cite{PlanckCollaborationVI2018}, where the middle 68th percentile is estimated to be $6.9 < z_{50\%} < 8.1$ with a $\tanh$-parametrisation of $\langle x_e \rangle$. 

\subsubsection{IGM temperature}
\label{sec:IGM_temperature}

In Table~\ref{tab:IGM_state} we also report the values of the temperature at the mean density, $T_0$. These are slightly higher than the log($T_0$/K)=4.21 $\pm$ 0.03 determined by \cite{Bolton2012} (the value is 3.9 $\pm$ 0.1 when correcting for the effect of \HeII ionization within QSO near zones).

It should be noted that the IGM temperature is very sensitive to the spectral shape adopted for the sources. While our approach avoids the introduction of additional degrees of freedom that comes with the assumption of a stellar spectrum (e.g. slope and shape), the results are strongly constrained by the hydrodynamic simulation\footnote{We have tested that adopting a simple power-law spectrum results in lower volume averaged temperatures $\vT$ and ionization fractions $\vHII$, whereas the neutral and partially ionized IGM is heated and ionized more.}.

\begin{figure}
\centering
\includegraphics[width=0.9\columnwidth]{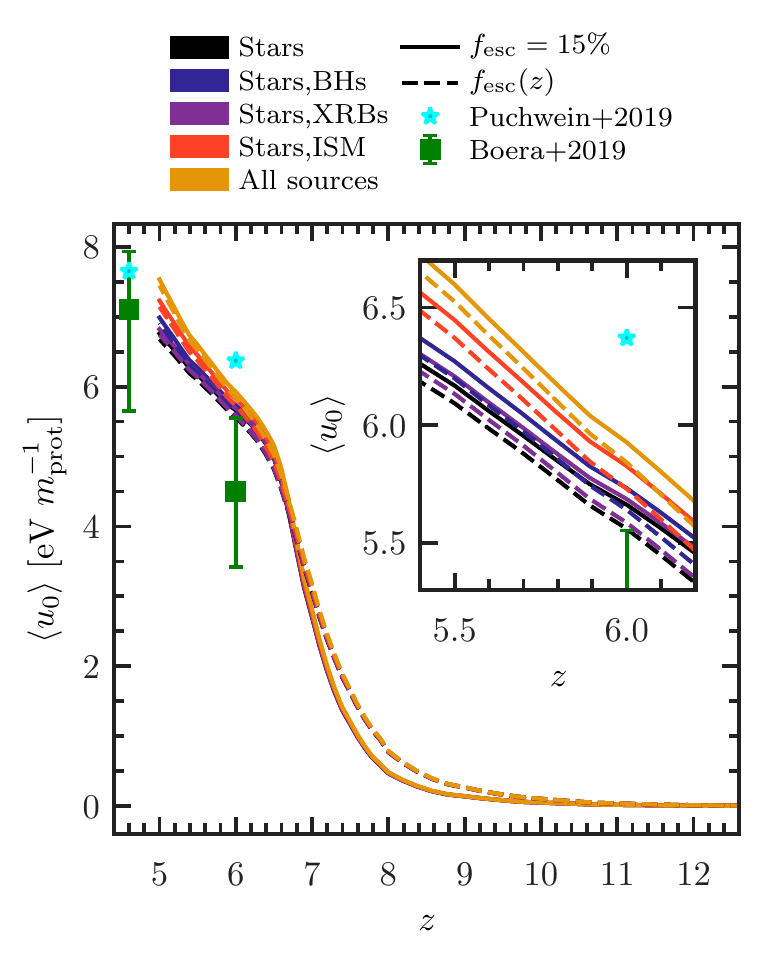}
\caption{Cumulative heating per unit mass, $\langle u_0 \rangle$, of simulations with different source types (indicated by the line colour---black: stars; blue: stars, BHs; purple: stars, XRBs; red: stars, ISM; yellow: all sources) and escape fraction (indicated by the line type--solid: 15\%; dashed: varying, $f_{\rm esc}(z)$). We also show observational constraints of \protect\cite{Boera2019} (green squares) and \protect\cite{Puchwein2019} (cyan stars).
}
\label{fig:u0}
\end{figure}
    
In Fig.~\ref{fig:u0} we show the cumulative energy deposited as heat per unit mass in the simulations, $\vun$, calculated following \cite{Boera2019} and \cite{Nasir2016}. This can be used as an independent constraint on reionization histories alongside the gas temperature. 
Independently from the source type, we observe a rapid increase in $\vun$, which roughly corresponds to the midpoint of reionization, $z_{50\%}$. This period of rapid heating is slightly shorter with a constant rather than a varying escape fraction, although in the latter case $\vun$ is $\sim 0.2$ eV~$m_{\rm prot}^{-1}$ larger at $z \sim 8$, consistently with the evolution of the gas temperature. The rate of heating declines at $z<6.6$ and it eventually becomes lower with $f_{\rm esc}(z)$ (see inset of the figure). At $z=6$ all our simulations produce values of $\vun$ in the range ($5.58-5.93$)~${\rm eV}\, m_{\rm prot}^{-1}$, higher than those derived by \cite{Boera2019} from measurements of the \Lya flux power spectrum, with the exception of models with stars only and stars plus BHs with an evolving $f_{\rm esc}$. On the other hand, our results at $z=6$ are all below the $\vun$ obtained by \cite{Boera2019} from the UV background (UVB) of \cite{Puchwein2019}, while at $z<6$, they seem to converge towards the \cite{Boera2019} and \cite{Puchwein2019} results at $z=4.6$.  Our $\vun$ histories show a more rapid increase in heating than \cite{Boera2019} find in their example models with the underlying \cite{Haardt2012} UVB. However, the $T_0$ they derive in their models and from their measurements ($T_0 = 7,600$~K), as well as the $T_0 = 12,000$~K of \cite{Puchwein2019}, are both lower than our values. The more rapid evolution of our $\vun$ and the higher $T_0$ can be understood as a consequence of the more rapid redshift evolution of the emissivity in our simulations, compared to those embedded in the UVB underlying the above estimates. Earlier ionization also gives ionized gas time to cool to a thermal asymptote. The majority of our IGM is reionized at later times (as seen in Fig.~\ref{fig:z_reion}) to temperatures similar to those expected by \cite{Boera2019} (who estimate the gas temperature of recently reionized gas to be 20,000~K) and has hence not undergone a similar cooling.

Our wide range of temperatures in the post-reionized regions is consistent with the thermal properties of the ionized gas in the full radiation-hydrodynamics simulation investigated by \cite{DAloisio2019}. 
As mentioned earlier, we find that the source spectral properties strongly affect also the thermal distribution, which is much more uniform in the presence of a power-law spectrum.
This dependence on the spectral shape is not found in \cite{DAloisio2019}, most probably because their spectra are cropped at $4$~Ryd. We note that, when employing a full RT, \cite{DAloisio2019} find temperatures higher than when using uniform UVB models (such as of~\citealt{Puchwein2019}), or RT simulations with low resolution or monofrequency spectra \citep[e.g.~][]{Keating2018}. This is consistent with test cases we have investigated. However, we do also find a wide range of temperatures on all scales in the post-ionization front zones (ranging from $\sim 18,000$~K to $\gtrsim 25,000$~K), highlighting that the simulated volume needs to be large enough for the results to be representative \citep{Iliev2014}. This does not rule out consistency between our findings and the lower $T$ in works with smaller scales, such as \cite{Finlator2018}. Our temperatures, though, are lower than those found by \cite{Ross2019} in their large scale RT simulations including QSOs and XRBs, where they predict $\vT \sim 200$~K at $z=14$, compared to the $\vT \sim 10$~K of our models. However, our approaches differ in the way galaxies are populated with QSOs (as they use N-body simulations) and the choice of the spectra, so that a detailed comparison is not straightforward. 

Finally, we note that the recent estimated constraints from the $21$~cm observations of LOFAR at $z \approx 9.1$ \citep{Ghara2020} find that, if the gas heating remains negligible, a mean ionization fraction of the IGM $\gtrsim 0.13$ is ruled out. These results align with the temperature inferences from PAPER-64 \citep{Greig2016} and SARAS 2 \citep{Singh2018}, which rule out a cold reionization scenario. Fig.~\ref{fig:T_neutral} suggests that the model including all source types is in closest  agreement with the LOFAR constraints.

\subsubsection{Thomson scattering optical depth}
\label{sec:optical_depth}
\begin{figure}
\includegraphics[width=\columnwidth]{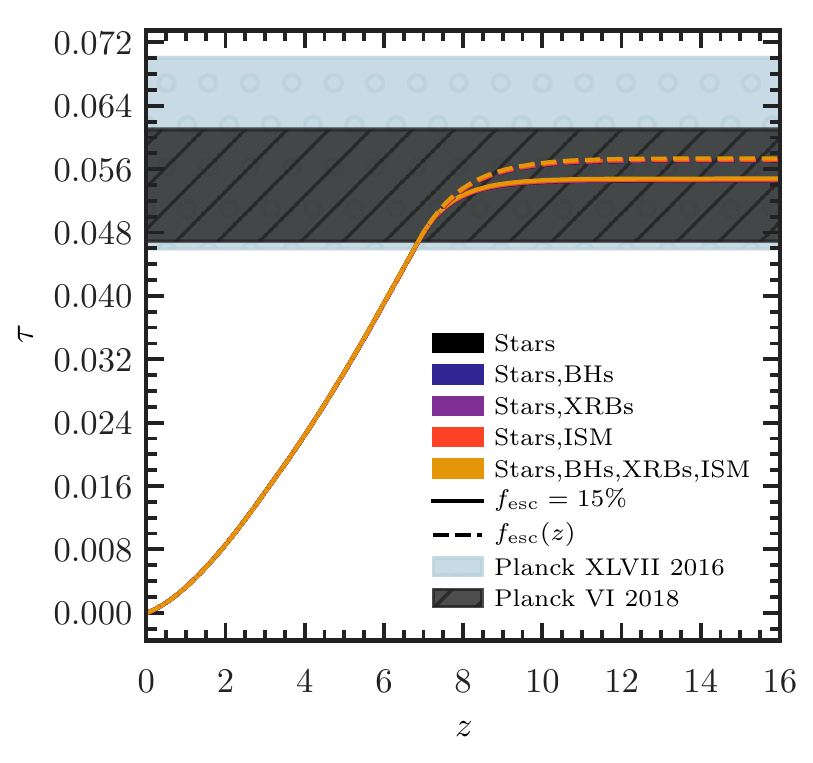}
\caption{Evolution of the Thomson scattering optical depth, $\tau$, for simulations with different source types (indicated by the line colour---black: stars; blue: stars, BHs; purple: stars, XRBs; red: stars, ISM; yellow: all sources) and escape fraction (indicated by the line type---solid: 15\%; dashed: varying, $f_{\rm esc}(z)$). We note that lines referring to simulations with the same escape fraction basically overlap. The dark dashed and blue circled regions refer to the constraints from~\protect\cite{PlanckCollaborationVI2018} and \protect\cite{PlanckCollaboration2016}, respectively. 
}
\label{fig:tau}
\end{figure}

In Fig.~\ref{fig:tau} we show the optical depth $\tau$ due to electron scattering for our simulations together with the $68\%$~CI optical depth $\tau = 0.054 \pm 0.007$ evaluated by the \cite{PlanckCollaborationVI2018} and the slightly larger $\tau = 0.058 \pm 0.012$ of \cite{PlanckCollaboration2016}.
The optical depth is calculated directly from the ionization fractions of our simulations at $z>5$, while we assume $x_i (3<z<5)=x_i(z=5)$ (where $i$= HII, HeII, HeIII) and that \HeII reionization ends instantaneously at $z_\HeII=3$, i.e.~that $\vHeIII = 1$ for $z \leq 3$\footnote{We evaluate $\tau$ also for $z_\HeII = 2.5$ and $3.5$ and find that the results are negligibly affected by this choice.}. 
Similarly to what we have observed for the reionization history, stellar type sources determine the optical depth, producing $\tau = 0.0546^{+0.0003}_{-0.0003}$ ($0.0571^{+0.0003}_{-0.0003}$ with a varying $f_{\rm esc}$), compared to the marginally larger $\tau = 0.0549^{+0.0003}_{-0.0002}$ ($0.0575^{+0.0003}_{-0.0002}$) obtained for all sources.
The simulations with a constant $f_{\rm esc}$ produce less ionization at high redshift, and consequently a lower $\tau$ than those with a varying $f_{\rm esc}$. We thus find that they are consistent with both the \cite{PlanckCollaborationVI2018} and the \cite{PlanckCollaboration2016} constraints.

\section{Discussion and Conclusions}
\label{sec:discussion}

\cite{Eide2018} found that during the Cosmic Dawn there is a subtle interplay between various sources of ionizing photons and that their collective impact on the reionization process is not simply given by a sum of the single components. This is particularly relevant for a correct determination of the IGM temperature, but also for partially ionized H and He ionization. Here we thus perform an analysis similar to \cite{Eide2018} concentrating on the Epoch of Reionization, at $z \lesssim 10$.
This is done by modelling the reionization of hydrogen and helium by post-processing the outputs of the SPH simulation MassiveBlack-II (\citealt{Khandai2015}) with the 3D radiative transfer code {\tt CRASH} (e.g. \citealt{Ciardi2001,Maselli2003,Graziani2013,Graziani2018}),
evaluating the physical properties of the IGM as determined by a variety of sources, namely stars, nuclear accreting black holes, X-ray binaries and shock heated ISM. The characteristics of the sources are derived directly from the properties of the stellar particles (such as stellar mass, age, metallicity for stars and XRBs), of the host galaxies (such as the star formation rate for the XRBs and the ISM) and of the BHs. Effectively, the only free parameter of the simulations is the escape fraction of UV photons.

Unlike most simulations in which the radiative transfer is coupled to the hydrodynamics (e.g. \citealt{Gnedin2014,OShea2015,Aubert2018,Rosdahl2018}), our simulations give results on cosmological scales large enough to provide representative predictions for the progression of reionization \citep{Iliev2014}, although they are still not large enough to capture the effect of the largest and rarest QSOs and cannot match the large scales (and parameter space) explored by semi-numerical  methods (e.g. \citealt{Mesinger2011,Fialkov2014,Hassan2017}). Being less computationally expensive, though, our  post-processing approach allows for a much higher sampling of the spectra of the sources in the radiative transfer, which is essential for a correct evaluation of the helium ionization and gas temperature in the presence of sources more energetic than stars. On the other hand, it fails to capture the feedback on structure formation from reionization itself. Our approach to post-processing differs in respect to those of other authors (e.g. \citealt{Ross2017,Keating2018}) as we use an SPH environment where the emissivities are not derived from dark matter halo properties, but rather from the properties of the sources as given by the hydrodynamic simulations (in essence, the only free parameter in our approach is the escape fraction of UV photons), in addition to having a very fine sampling of their spectrum.  

Our results can be summarized as follows.
\begin{itemize}
\item Consistently with previous works and with Eide2018, full hydrogen reionization is driven by stars, which create rapidly growing and overlapping bubbles. The volume average \HII fraction at $z\sim 6$ is in fact $0.99998$ in all the scenarios investigated.
\item Photons from more energetic sources (XRBs and ISM) propagate further into the IGM, inducing partial ionization and heating of the gas, in both its hydrogen and helium components, with values $\xHII \sim \xHeII \sim 10^{-4}-10^{-2}$ and temperatures as high as $\sim 10^3$~K. Due to its soft spectrum, the ISM is particularly efficient. These results are consistent with what discussed by Eide2018 at higher redshift.
\item While BHs do not have a strong impact on the average properties of the IGM due to their paucity, whenever active they dominate the local production of ionizing photons, increasing the extent of the \HII regions. More importantly, they are the only sources capable of fully ionize helium (as it can be only partially ionized by XRBs and ISM), increasing the volume average \HeIII fraction at $z \sim 6$ to $\sim 0.02$ compared to $\sim 0.0002$ with stars only, $0.0009$ with XRBs and $0.01$ with ISM. 
\item In the vicinity of energetic sources, where values $\xHeIII \gtrsim 10^{-2}$ are reached, the gas temperature increases as much as $10^4$~K compared to the case in which only radiation from stars is considered. This confirms the importance of including the treatment of the helium component of the gas for a correct evaluation of the temperature. Similarly, because both the ionization level of helium and the value of the temperature are highly dependent on the spectral shape of the ionization field, it is equally crucial to sample the spectrum with a high accuracy.
\item Among the models explored here, we find that simulations with an escape fraction that decreases with decreasing redshift reproduce more closely observational data, including the volume averaged emissivity. This also suggests that a constant value lower than the 15\% adopted here could give a better fit. It should be noted, though, that the escape fraction is degenerate with properties such as the SFR and the stellar IMF. These are determined by the hydrodynamic simulations and the only free parameter thus remains $f_{\rm esc}$.
\item Our models are consistent with constraints on the Thomson scattering optical depth from Planck, as well as on the volume average \HI fraction down to $z \sim 6$, while a sub-grid treatment of the LLSs is necessary for a better modelling of the properties of the IGM at lower redshifts.
\item Finally, our thermal histories are all dominated by the \HII ionization of stars, reaching mean gas temperatures of $\vT = 19,274$~K ($18,643$~K with an evolving escape fraction) with stars alone at $z=6$. These temperatures only increase by at most $\sim 1,000$~K when including other source types. At $z<6$ we see further heating from \HeII ionization. Our temperatures are marginally larger than those deduced from observations, such as the temperature at mean density $T_0$ and the cumulative heating per unit mass $\vun$, and we conclude that this mainly is a feature of the redshift evolution of the ionizing emissivity. Important to the \HI 21~cm line, the thermal state of the neutral or lowly ionized IGM is extremely sensitive to the presence and abundance of energetic sources. With only stars or additionally the BHs, we expect a 21~cm signal in absorption even at $z<10$, whereas we expect a strong (XRBs) and stronger (ISM) signal in emission with the other source types. Observations (PAPER-64, SARAS 2, and LOFAR) favour our model with all source types combined.
\end{itemize}

Although clear differences in the IGM properties emerge depending on the source combinations, the observational data used in this paper are not able to unambiguously discriminate between different source types and their relative contribution to the reionization process. Additional information is expected from the 21~cm line from neutral hydrogen (Ma et al. in prep), but, to maximize the extraction of information, it will be crucial to cross-correlate this signal with observations in different frequency bands
such as of \Lya \citep[e.g.~][]{Vrbanec2020} or  \OIII \citep{Moriwaki2019} emitters, or of integrated quantities as X-ray \citep[e.g.~][]{Ma2018} and infrared \citep[e.g.~][]{Fernandez2014} background, and the CMB \citep[e.g.~][]{Jelic2010,Ma2018b}.

Whereas redshifted \OIII may be observed with JWST, two other hyperfine transitions holds the potential to be probes of the high-$z$ IGM \citep[e.g.~][]{Furlanetto2006}. The $3.46$~cm transition of \threeHplus can be a unique signature of \HeIII regions \citep{Khullar2020}, which we have found to overwhelmingly exist only in the vicinity of BHs. Furthermore, this \threeHplus signal and the even more elusive \DI $91.6$~cm transition of deuterium may be observables less prone to the foreground contamination that occludes \HI 21 cm signals.

To conclude, our large scale radiative transfer simulations of IGM reionization, anchored to detailed hydrodynamic simulations, give a new insight into the relative role of a variety of source types. The exceptional spectral resolution employed, as well as the inclusion of the helium component of the gas, also assure an accurate evaluation of the IGM temperature.

\section*{Acknowledgements}
We thank Piero Madau, Martin Glatzle, Enrico Garaldi, Joakim Rosdahl and Max Gronke for enlightening discussions, and an anonymous referee for insightful comments that helped to decisively improve the manuscript. MBE is grateful to the Astronomy and Astrophysics department at UCSC and the Institute of Theoretical Astrophysics at UiO for their hospitality. The stay at UCSC was funded through a U.S.-Norway Fulbright scholarship. LG acknowledges support from the Amaldi Research Center funded by the MIUR program "Dipartimento di Eccellenza" 
(CUP:B81I18001170001).

We are grateful for the availability of open source software. In this work, we made use of \texttt{numpy} \citep{VanderWalt2011}, \texttt{cython} \citep{Behnel2011}, \texttt{Yt} \citep{Turk2011} and \texttt{Matplotlib} \citep{Hunter2007}.

\section*{Data availability}

No new data were generated or analysed in support of this research.


\bibliographystyle{mnras} 
\bibliography{main} 

\appendix

\section{Loss of high energy photons}
\label{app:loss}
While energetic photons are important for partial ionization and heating, their mean free path is very long and thus, with our assumption of non-periodicity in the RT, they could easily escape from the simulation box and be lost\footnote{We do however account for and track the redshifting of each photon as it propagates through the box.}. In the following we show that this is not expected to have a significant impact on the results presented in this paper.

\begin{figure}
    \centering
    \includegraphics{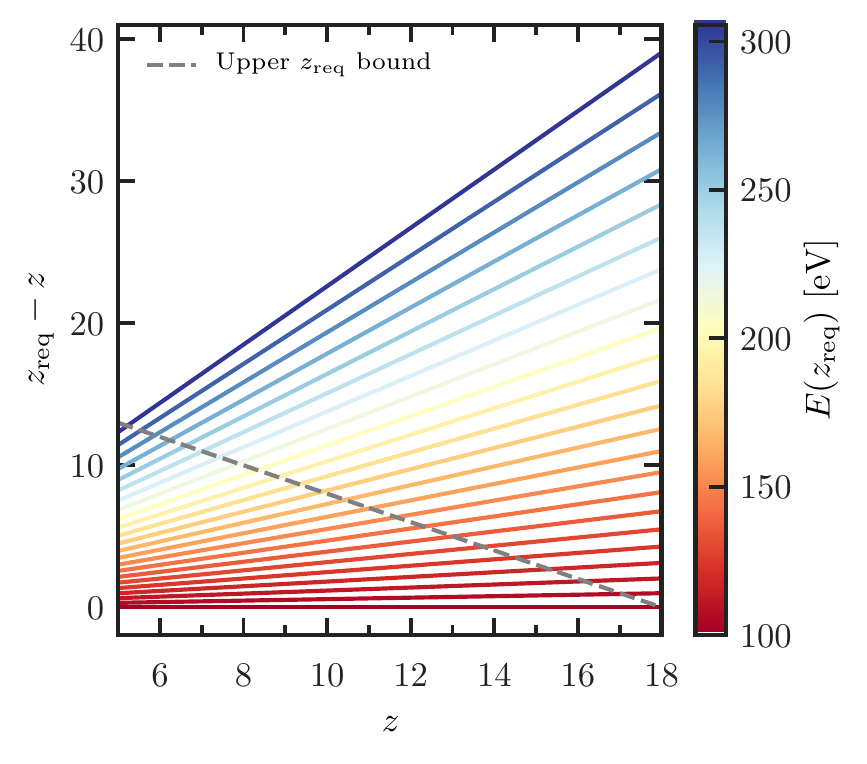}
    \caption{Redshift at which a photon with energy indicated by the color bar needs to be emitted in order to be shifted to 100~eV by redshift $z$ indicated in the x-axis. Photons occupying the area above the grey dashed line would need to be emitted at redshifts above those covered by our simulations.}
    \label{fig:redshift}
\end{figure}

In Fig.~\ref{fig:redshift} we plot at which redshift, $z_{\rm req}$, a photon of energy $E$ needs to be emitted in order to be redshifted to 100~eV (and thus be easily absorbed) by a given redshift $z$. As a reference, to reach 100~eV by the end of the simulation at $z\sim 5$, a photon of 300~eV should have been emitted at $z \sim 17$. As the IGM has already been partially or fully ionized well before $z=5$ by less energetic photons, this suggests that photons above a couple of hundred eV need times longer than those available in the simulation to possibly play a relevant role. 

\begin{figure}
    \centering
    \includegraphics{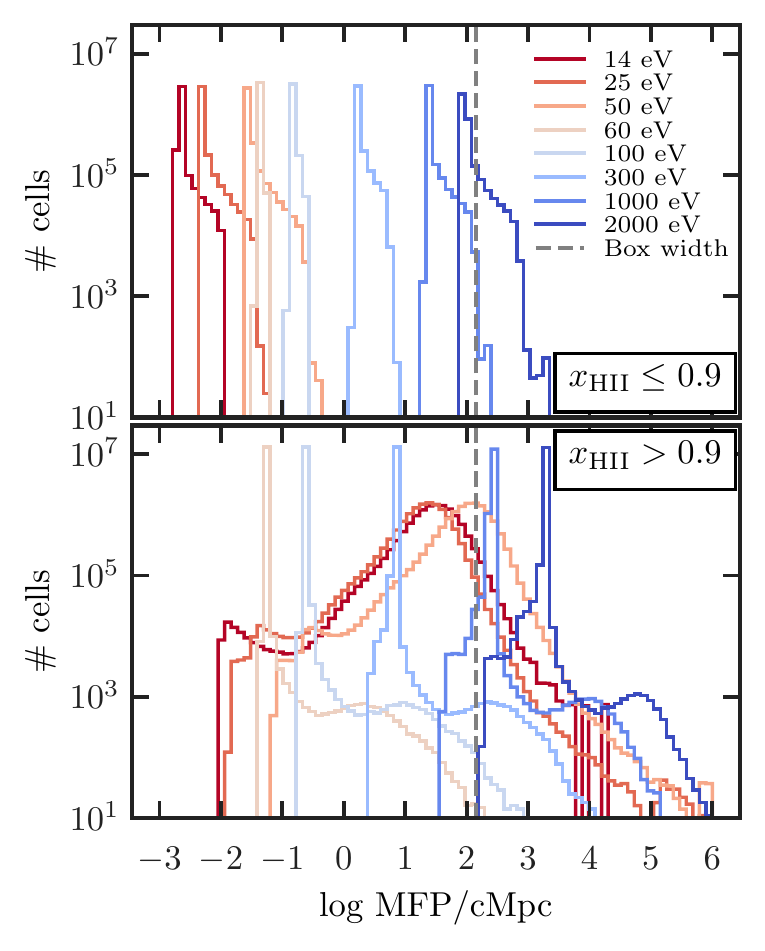}
    \caption{Number of cells with a given mean free path in the simulation with all source types and $f_{\rm esc}(z)$ at $z=7$. The distributions refer to photons of different energies (as indicated by the colors), while the vertical grey dashed line indicates the box dimension. The top and bottom panels refer to cells with $\xHII \leq 0.9$ and $\xHII > 0.9$, respectively.}
    \label{fig:mfpcells}
\end{figure}

In Fig.~\ref{fig:mfpcells} we plot the distribution of the mean free path (MFP) of cells in the simulation with all source types and $f_{\rm esc}(z)$ at $z=7$. When only highly neutral cells (i.e. with $\xHII \leq 0.9$) are considered (top panel), the distributions are strongly peaked and shift to larger MFP with increasing photon energy. For all photons with $h_{\rm P}\nu < 100$~eV the MFP is smaller than the box dimension. When only highly ionized cells (i.e. with $\xHII >0.9$) are considered (bottom panel), instead, the distributions are more complex, as the presence of He (and its ionization state) becomes more relevant. We then see that the MFP of photons below the ionization threshold of \HeII becomes even larger than the box size, as \HI and \HeI are almost fully ionized, and the corresponding distributions are much wider, reflecting the strong dependence on the detailed ionization state of the various components of the gas. Conversely, higher energy photons have distributions which still resemble those in the left panel, with a strong peak, albeit shifted towards larger values of the MFP due to the reduced absorption from \HI and \HeI. 

\begin{figure}
    \centering
    \includegraphics[width=0.85\columnwidth]{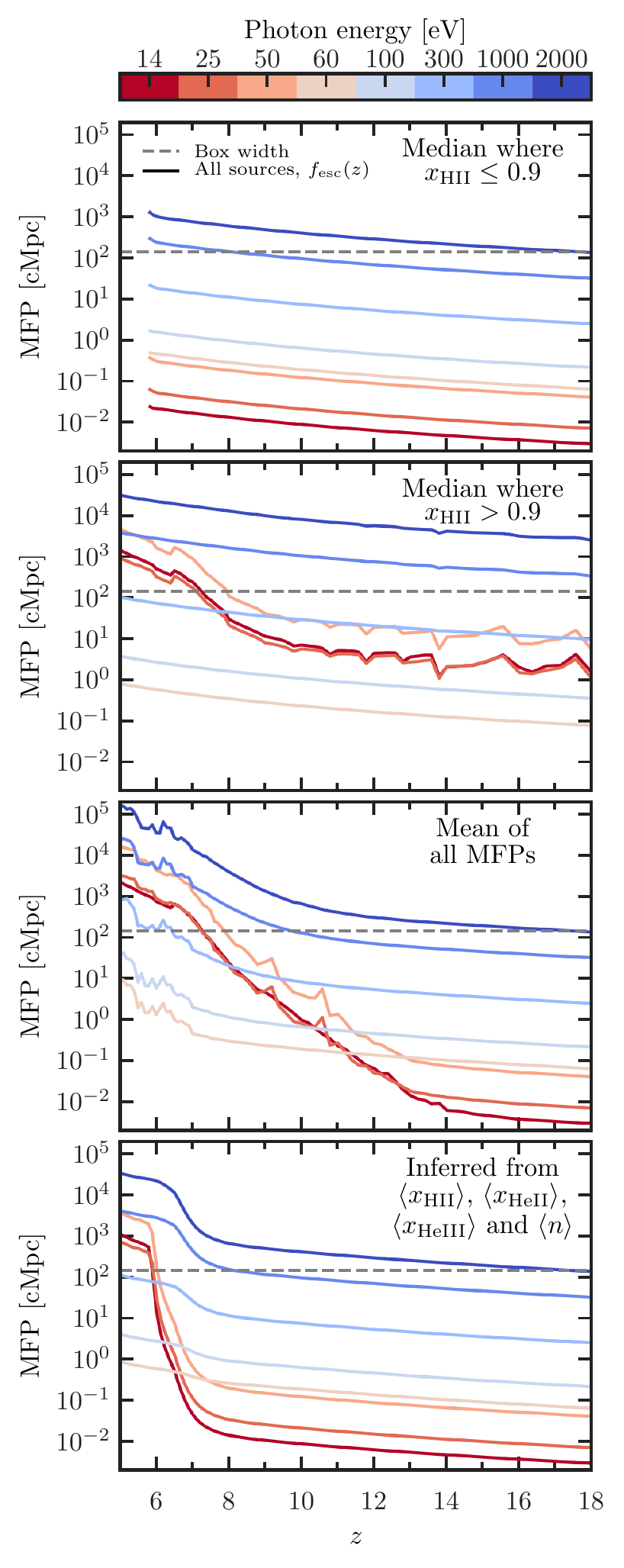}
    \caption{Redshift evolution of the mean free path (derived from the statistics of the MFPs in each simulation cell) of photons of different energies (as indicated by the line colour) in the simulation with all source types and $f_{\rm esc}(z)$. The grey dashed lines refer to the length of the simulation box. From top to bottom the panels refer to: the median of the MFPs in cells with $\xHII \leq 0.9$; the median in cells with $\xHII > 0.9$; the volume average of all the MFPs; and the MFPs one would obtain by calculating them directly from the volume averaged number density $\langle n \rangle$ and ionization fractions \vHII, \vHeII{} and \vHeIII of the simulation.}
    \label{fig:mfp}
\end{figure}

In Fig.~\ref{fig:mfp} we show the redshift evolution of the MFPs for photons of different energies. These have been calculated by taking the median of the MFPs in cells where hydrogen is highly neutral ($\xHII \leq 0.9$, first panel) as well as where it is highly ionized ($\xHII > 0.9$, second panel); by taking the mean of the MFPs in all cells (third panel) and finally by inferring the MFP from the volume averaged ionization fractions $\vHII$, $\vHeII$ and $\vHeIII$. 
The results in the highly neutral IGM reflect what was observed in the previous figure, i.e.~the MFP smoothly increases with increasing photon energy. All photons with energies below a couple of hundred eV have MFPs shorter than the box size at all redshifts. Conversely, for the highest energy photons this happens only at high redshift. For example, the MFP of a 1 keV photon becomes larger than $100 h^{-1}$ cMpc at $z<8$. 
This increase in MFP with decreasing redshift observed for all photon energies is associated to the declining gas number density with cosmological expansion and is proportional to $1/z^3$. 

In highly ionized regions (second panel), the MFP of photons with energies below 54.4~eV becomes much longer due to the almost lack of \HI and \HeI, and once reionization is well under way it becomes even larger than the box size. For these photons there is also a stronger dependence on redshift, with a more rapid increase observed from $z \sim 9$, when $\xHII > 10^{-1}$. 
Photons with energies above 54.4~eV, instead, have a behaviour similar to the one observed in the previous plot, but the MFPs can be more than one order of magnitude larger. This difference means that even the highest energy photons are sensitive to the presence (or absence) of \HI (and \HeI), even though the ionizing cross section of hydrogen $\sigma_\HI$ at such energies is extremely small (as $\sigma_\HI \propto (h_P \nu /{13.6\, \rm eV})^{-3}$). 

The volume averaged MFP (irrespective of \HI ionization state, as plotted in the third panel) displays how the MFPs transition from being dominated by those found in the neutral \HI at high $z$, to the much longer ones in \HII regions at lower $z$. This increase is most dramatic and extended for \HI and \HeI ionizing photons, whose MFPs increase by more than five orders of magnitude between $z\sim14$ and the end of the simulation. For higher energy \HeII ionizing photons we see the same uptick in MFPs at $z<6$, when \HeII ionization becomes significant. 

Finally, in the last panel we show the MFPs one would obtain by calculating them directly from the volume averaged ionization fractions $\vHII$, $\vHeII$ and $\vHeIII$ and number density. The MFPs display the same behaviour as in the third panel, albeit with a much shorter transition period at $z<7$ from being dominated by MFPs in the neutral \HI to the ionized \HII  at $z=6$. 

We conclude by observing that even very high energy photons have MFPs similar to our box length, as long as some \HI is still present. It should be highlighted, though, that photons above a couple of hundred eV need times longer than those available in the simulation to possibly play a relevant role.

\section{Reionization timing}
\label{app:reionization_timing}

\begin{figure}
\centering
\includegraphics[width=0.9\columnwidth]{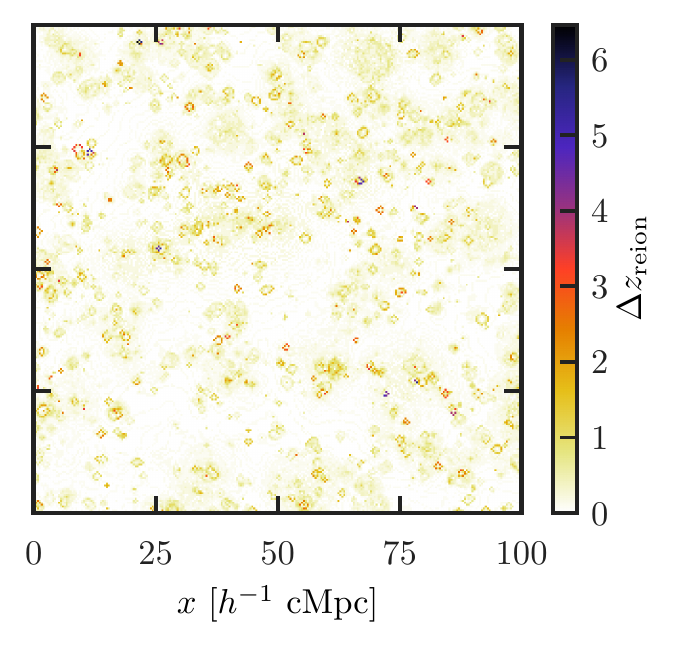}
\caption{Map showing the difference in the timing of reionization, $z_{\rm reion}$, for a simulation with all source types and $f_{\rm esc} = 15\%$ when we adopt an ionization threshold $\xHII^{th}=0.1$ with respect to the case when the threshold is $0.9$. }
\label{fig:delta_z_reion_threshold}
\end{figure}

We define the reionization redshift $z_{\rm reion}$ of the cell as the redshift at which the cell has reached a hydrogen ionization fraction $\xHII$ larger than a threshold value, $\xHII^{th}$. In our reference case we adopt $\xHII^{th} = 0.9$.
In Fig.~\ref{fig:delta_z_reion_threshold} we show how $z_{\rm reion}$ changes when we lower this threshold 0.1 for the simulation with all source types and $f_{\rm esc} = 15\%$. We find that $z_{\rm reion}$ increases by as much as $\Delta z_{\rm reion} \sim 2$ in the vicinity of the sources, while it remains unaltered for the majority of the IGM, with $\Delta z_{\rm reion} < 0.1$ for $50\%$ of the gas, and $\Delta z_{\rm reion} < 0.4$ for $90\%$ of the gas.

\begin{figure}
\includegraphics[width=0.9\columnwidth]{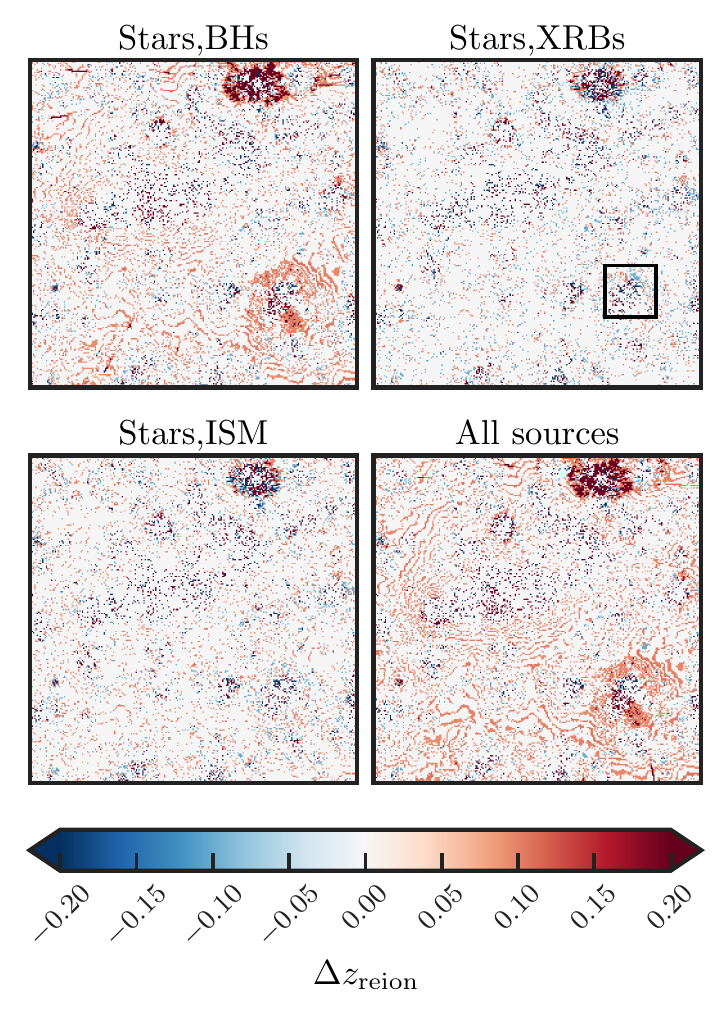}
\caption{Maps showing the differences in the timing of reionization, $z_{\rm reion}$, for simulations with different source types (as indicated by the labels), $f_{\rm esc} = 15\%$ and an ionization threshold $\xHII^{th}=0.9$, with respect to the simulation with only stars. The maps are $100 h^{-1}$ cMpc wide. The black square highlights a region in which the features resembling Monte Carlo noise are actually due to physical reasons (see text for more details).}
\label{fig:delta_z_reion}
\end{figure}

In Fig.~\ref{fig:delta_z_reion} we show how $z_{\rm reion}$ changes with different source types for a reference value of $\xHII^{th}=0.9$. We find that reionization in the vicinity of BHs happens earlier, by more than a factor of 0.2 in redshift, whereas
XRBs and ISM have a very small impact. Although some of the features observed in all these panels closely resemble Monte Carlo noise, we have verified that this is not the case for all cells and they are instead related to the complexity of the multi-frequency radiative transfer. We highlight one such region with a square in the `Stars,XRBs' panel of the figure. Indeed some gas pockets experience earlier reionization due to the longer mean free path of the high energy photons emitted by these sources, but at the same time, because such photons ionize less efficiently, for some cells we observe a delayed reionization.
The behaviour of the simulation including all source types reflexes that observed in the simulation with stars and BHs, with reionization occurring slightly earlier ($\Delta z_{\rm reion} \approx 0.1$) in the vicinity of BHs.

\bsp	
\label{lastpage}
\end{document}